\documentclass[twocolumn]{aastex62}

\usepackage{amsmath}
\usepackage{amssymb,bm}

\graphicspath{{./}{figures/}}

\received{XXX}
\revised{YYY}
\accepted{ZZZ}

\submitjournal{ApJ}

\newcommand{\be}{\begin{equation}}
\newcommand{\ee}{\end{equation}}
\newcommand{\bea}{\begin{eqnarray}}
\newcommand{\eea}{\end{eqnarray}}
\def\ba#1\ea{\begin{align}#1\end{align}}

\def\({\left(}
\def\){\right)}

\def\[{\left[}
\def\]{\right]}

\newcommand{\refeq}[1]{Eq.~(\ref{eq:#1})}

\newcommand{\reffig}[1]{Fig.~\ref{fig:#1}}
\newcommand{\refsec}[1]{Sec.~\ref{sec:#1}}
\newcommand{\refapp}[1]{App.~\ref{app:#1}}

\newcommand{\vs}{\nonumber\\}

\newcommand{\plotthree}[3]{\includegraphics[width=.31\textwidth]{#1}
\includegraphics[width=.31\textwidth]{#2} \includegraphics[width=.31\textwidth]{#3}}

\def\vy{\bm{y}}
\def\vz{\bm{z}}

\def\vx{\bm{x}}
\def\vq{\bm{q}}
\def\bfx{\bm{x}}

\def\vk{{\bm{k}}}

\def\bfq{\bm{q}}

\shorttitle{Parallel polyspectra estimator}
\shortauthors{Tomlinson et al.}

\definecolor{RedWine}{rgb}{0.743,0,0}
\definecolor{GrassGreen}{rgb}{0.125,0.75,0.125}
\definecolor{RoyalBlue}{rgb}{0.25,0.41,0.88}

\begin{document}

\title{Efficient parallel algorithm for estimating higher-order polyspectra}

\correspondingauthor{Donghui Jeong}
\email{djeong@psu.edu}

\author[0000-0001-8207-000X]{Joseph Tomlinson}
\affil{Department of Astronomy and Astrophysics and Institute for Gravitation and the Cosmos, \\
The Pennsylvania State University, University Park, PA 16802, USA}

\author[0000-0002-8434-979X]{Donghui Jeong}
\affil{Department of Astronomy and Astrophysics and Institute for Gravitation and the Cosmos, \\
The Pennsylvania State University, University Park, PA 16802, USA}

\author[0000-0002-4391-2275]{Juhan Kim}
\affiliation{
Center for Advance Computation, Korea Institute for Advanced Study, 85 Hoegiro, Dongdaemun-gu, Seoul 130-722, Korea}

\begin{abstract}
Nonlinearities in the gravitational evolution, galaxy bias, and redshift-space distortion drive the observed galaxy density fields away from the initial near-Gaussian states. Exploiting such a non-Gaussian galaxy density field requires measuring higher-order correlation functions, or, its Fourier counterpart, polyspectra. Here, we present an efficient parallel algorithm for estimating higher-order polyspectra.  Based upon the Scoccimarro estimator, the estimator avoids direct sampling of polygons by using the Fast-Fourier Transform (FFT), and the parallelization overcomes the large memory requirement of the original estimator. In particular, we design the memory layout to minimize the inter-CPU communications, which excels in the code performance.
\end{abstract}

\keywords{
cosmology: large-scale structure of universe --- methods: data analysis
}

\section{Introduction} \label{sec:intro}
The primordial density field in the Universe measured by the temperature anisotropies and polarizations of cosmic microwave background (CMB) radiation is very close to Gaussian \citep{Planck2015:fnl}. In contrast, the observed galaxy density field shows strong non-Gaussianity due to late-time effects such as nonlinear gravitational instability, nonlinear galaxy bias as well as nonlinear redshift-space distortion.

The analysis of the galaxy density field in literature is yet focusing on the galaxy two-point correlation function, or its Fourier counterpart, the galaxy power spectrum \citep{2006PhRvD..74l3507T, 2011MNRAS.415.2876B,2012PhRvD..86j3518P,2012MNRAS.426.2719R,2013A&A...557A..54D,2017A&A...604A..33P}). At the same time, the interest in the higher-order correlation functions and polyspectra has emerged to complement the two-point statistics in exploiting such a non-Gaussian galaxy density field. In Fourier space, these are, from third order, bispectrum, trispectrum, quadspectrum, pentaspectrum, hexaspectrum, and so on.

Thus far, the analysis on the higher-order correlation functions and polyspectra focuses mainly on the galaxy bispectrum and three-point correlation function \citep{2001ApJ...546..652S,2002MNRAS.335..432V,2005MNRAS.364..620G,2006MNRAS.368.1507N,2007PASJ...59...93N,2011ApJ...737...97M,2011ApJ...739...85M,2011ApJ...726...13M,2015MNRAS.451..539G,2013MNRAS.432.2654M}, but they have not been as informative as their lower-order counterparts. This is due to both the high computational cost in estimating higher-order correlation functions and the lack of accurate modeling needed for detailed data analysis.

One challenge in studying the higher-order polyspectrum in Fourier space is to find an efficient estimator; when estimating $n$-th order polyspectrum, the complexity of the naive estimator grows as the number of all possible $n$-gons in the three-dimensional Fourier space that scales as $N_{\rm max}^{3(n-1)}$. Here, $N_{\rm max}\equiv k_{\rm max}/\delta k$ is the number of one-dimensional discrete grid points of interest, with $k_{\rm max}$ being the maximum wavenumber of the analysis, usually set by the smallest scale that we can reliably model the nonlinearities in galaxy density field, and $\delta k$ being the size of the binning in Fourier space. 

One can alleviate this computational cost by using the Fourier space representation of the Dirac-delta operator that dictates the statistical homogeneity. Originally done by \citet{scoccimarro:2000} for estimating the bispectrum, in this way, we can estimate a polyspectrum by taking advantage of the Fast Fourier transformation algorithm. As we shall outline in \refsec{serial}, the calculation of the polyspectrum then boils down to a sum of products of the $N_{\rm max}$ distinct three-dimensional Fourier volumes derived from the original density field. This method avoids directly sampling all $n$-gons and is computationally much faster for all high-order polyspectra.

The implementation of the Fourier based method, however, requires allocating a large amount of memory. To avoid the aliasing problem associated with the Fourier method, the three-dimensional Fourier grid must be at least $(sN_{\rm max})^3$ for the $s$-point polyspectrum; as we need a total of $N_{\rm max}$ such three-dimensional Fourier volumes, the memory requirement grows as $s^3N_{\rm max}^4$. For example, the memory requirement to estimate the bispectrum ($s=3$) with $N_{\rm max}=128$, when using single precision, is already 27 Gigabytes.

In this paper, we shall improve the performance of the polyspectra estimators by parallelizing them. The most straightforward and naive approach might be to distribute each of the Fourier volumes to a different processor. To compute the polyspectum, then we need to calculate the sum of the multiplication of the Fourier volumes in different processors. This proves to be quite slow and, if not slower, only marginally faster than even the direct calculation from sampling all possible $n$-polygons. Instead, we distribute the Fourier volumes to parallel nodes so that the summation can be done almost entirely locally. Here, most of the inter-processor communication occurs during the Fourier transform step and the performance is significantly faster than the serial estimator. In this paper, we shall demonstrate the performance of this parallel algorithm for efficient calculation of higher-order polyspectra. 

This paper is organized as follows. In \refsec{serial} we discuss the Scoccimarro estimator and its extension to higher-order polyspectra and the limitations of such an extension. In \refsec{parallel} we explain our efficient method of parallelization for this estimator. \refsec{calcnk} analyzes the results of our application of the estimator to a grid of unity, equivalent to calculating the number of polygons, for a variety of high-order polyspectra and also includes an analysis of the scaling of our parallel algorithm. We conclude in \refsec{conclusion}. \refapp{sum} discusses the conversion of our general estimator from integral to summation form for direct application to calculations. Finally, \refapp{analytic} discusses some approximately analytic solutions to the number of $n$-gons in three-dimensional Fourier space.

We use the following convention 
\ba
\delta(\vk)=&\int d^3x \delta(\vx) e^{-i\vk\cdot\vx}
\,,
\vs
\delta(\vx)=&\int \frac{d^3k}{(2\pi)^3}
\delta(\vk) e^{i\vk\cdot\vx}
\,,
\ea
for the Fourier transformation, and 
\be
\left<
\delta(\vk_1)\cdots\delta(\vk_n)
\right>
=
(2\pi)^3P_{n}(\vk_1,\cdots,\vk_n)
\delta^D(\vk_{1\cdots n})
\ee
for the $n$-point polyspectrum. 
Here, $\delta^D$ is the Dirac-delta operator, and we use the shorthanded notation 
of $\vk_{1\cdots n}\equiv \vk_1+\cdots\vk_n$. 
We denote the amplitude of a vector $\vk_i$ as $k_i$.

\section{The Scoccimarro estimator} \label{sec:serial}
The estimator for polyspectra that we shall parallelize here is based on the Scoccimarro estimator \citep{scoccimarro:2000,scoccimarro:2015} that we summarize in this section. In this section, we shall mainly discuss the algorithm with the equations in the continuous limit and summarize the discrete implementation in \refapp{sum}.

Let us begin with the bispectrum estimator. The estimator for the monopole bispectrum $B(k_1,k_2,k_3)$ takes the average of $\delta(\vq_1)\delta(\vq_2)\delta(\vq_3)$ for all possible combinations of three vectors $|q_i-k_i|<\delta k_i/2$ ($i=1,2,3$) that satisfy the triangle condition ($\vq_{123}=0$). We denote the bin size in $i$-th direction as $\delta k_i$, although, throughout the paper, we choose the bin size for all three directions the same for simplicity. The extension of the method to arbitrary bin size is trivial. We may formulate the bispectrum estimator as
\ba
B(k_1,k_2,k_3)
=&
\frac{1}{VN_{(123)}}
\sum_{q_1\sim k_1}
\sum_{q_2\sim k_2}
\sum_{q_3\sim k_3}
\vs
&\times
\delta(\vq_1)
\delta(\vq_2)
\delta(\vq_3) \delta^D(\vq_{123})\,,
\label{eq:estimateBk}
\ea
with $N_{(123)}$ being the number of triplets 
$(\vq_1,\vq_2,\vq_3)$ satisfying the binning condition 
($|q_i-k_i|<\delta k_i$ which is denoted as $q_i\sim k_i$)
and the triangle condition:
$
N_{(123)} = \sum_{\vq_1,\vq_2,\vq_3} \delta^D(\vq_{123})
$
with the same summation as \refeq{estimateBk}. Throughout the paper, we impose the condition that $k_1\ge k_2\ge k_3$ to avoid the duplication.

In the continuous limit, the bispectrum estimator becomes (see, \refapp{sum} for the details about the normalization factor)
\ba
B(k_1, k_2,k_3) 
=&
\frac{V_f}{V^B_{(123)}(2\pi)^3}\int_{k_1}d^3q_1\int_{k_2}d^3q_2\int_{k_3}d^3q_3
\vs
&\times
\delta(\vq_1)\delta(\vq_2)\delta(\vq_3) \delta_D(\mathbf{q}_{123})
\label{eq:Bkdef}
\ea
where $V_f$ is the volume of a fundamental Fourier cell, $(2\pi)^3/V$, and 
$V^B_{(123)}$ is the Fourier space volume defined as 
\begin{eqnarray}
&V^B_{(123)}& = \int_{k_1}d^3q_1\int_{k_2}d^3q_2\int_{k_3}d^3q_3
\delta_D(\mathbf{q}_{123})\,.
\end{eqnarray}
The key step for the Scoccimarro estimator is promoting the Dirac-delta operator with its Fourier representation:
\be
\delta_D(\vq_{123})
=
\int\frac{d^3x}{(2\pi)^3} e^{-i\vx\cdot(\vq_1+\vq_2+\vq_3)}\,,
\ee
with which we can recast the three integrals for sampling triangles in \refeq{Bkdef} into single integration:
\be
B(k_1, k_2,k_3) 
=
\frac{V_f}{V^B_{(123)}(2\pi)^3}
\int \frac{d^3x}{(2\pi)^3} 
I_{k_1}(\vx)
I_{k_2}(\vx)
I_{k_3}(\vx)\,.
\label{eq:Bkfast}
\ee
Here,
\be
I_{k_i}(\vx)
\equiv
\int_{k_i} d^3q \delta(\vq) e^{-i\vq\cdot\vx}
\equiv
\int d^3q 
\tilde{I}_{k_i}(\vq) e^{-i\vq\cdot\vx}
\,,
\ee
is the Fourier transformation of the function $\tilde{I}_{k_i}(\vq)$ that is defined with the density field in the spherical shell defined by $|q - k_i|<\delta k_i/2$ and zero otherwise.
That is, instead of time-consuming operation of sampling all possible triangles, the Scoccimarro estimator involves Fourier transformation for computing $I_{k_i}(\vx)$ at each wavenumber bin, and summing over the 
product of three $I_{k_i}$s. 

Note that the complexity, or the scaling of the computation time as a function of system size, of the Scoccimarro estimator is the same as the direct summation's scaling of $N_{\rm max}^{6}$ for the case of bispectrum. The Scoccimarro estimator, however, reduces the computation time in two ways. First, we have fewer total numerical operations. The direct sampling involves looping over the full ranges of 
$-N_{\rm max}\sim N_{\rm max}$
for each component of $\vk_1$ and 
$\pm N_{\rm max}/2\sim \pm N_{\rm max}$
for each component of $\vk_2$, because our convention of $k_1\ge k_2\ge k_3$ with triangle condition constrains $k_1/2\le k_2\le k_1$. On the other hand, the Scoccimarro estimator reduces it to the matrix inner product ($N_{\rm max}^3$ operation) only for the amplitude triplets $(k_1,k_2,k_3)$. Second, for the Scoccimarro estimator, the three-dimensional array $\delta(\vq)$ is accessed in an ordered manner that is easier for the CPU to cache. In contrast, the direct sampling accesses $\delta(\vq_3)$ in an irregular manner, which is inevitable because the value for $\vq_3$ is determined by $\vq_3 = -\vq_{1}-\vq_2$. Combining these two factors, the serial version of the Scoccimarro estimator is a factor of 10 faster than the direct sampling in our implementation.

The reduction of operation time is even more striking as we calculate the 
higher order correlation functions. When computing the `angle-averaged' 
$n$-point polyspectra that are only a function of the Fourier 
wavenumbers \citep{2005PhDT........23S}, the Scoccimarro estimator's complexity
increases only by one more power of $N_{\rm max}$, 
${\cal O}(N_{\rm max}^{n+3})$ while the direct 
sampling method's complexity grows by three powers of $N_{\rm max}$ at
each order: ${\cal O}(N_{\rm max}^{3(n-1)})$. 
Needless to say that the base operation time for Scoccimarro estimator is much faster than the direct sampling case, as we have already seen in the case of the bispectrum where the two method have the same complexity.

Extending the bispectrum estimator in \refeq{Bkfast} for the angle 
averaged higher-order polyspectra is straightforward:
\ba 
&\tilde{P}_n(k_1\cdots,k_n) 
\vs
=& \frac{V_f}{V^{\tilde{n}}_{(12\cdots n)}(2\pi)^3} \int_{k_1}d^3q_1\cdots\int_{k_n}d^3q_n \delta_D(\mathbf{q}_{1\cdots n})\prod \delta(\mathbf{q}_i) \nonumber
\vs
=& 
\frac{V_f}{V^{\tilde{n}}_{(12\cdots n)}(2\pi)^3} \int
\frac{d^3x}{(2\pi)^3}
I_{k_1}(\vx)\cdots I_{k_n}(\vx)
\label{eq:estimator}
\ea
with the Fourier volume
\begin{eqnarray} \label{eq:volume}
    V^{\tilde{n}}_{(12\cdots n)} = \int_{k_1}d^3q_1\cdots\int_{k_n}d^3q_n \delta_D(\mathbf{q}_{1\cdots n})\,.
\end{eqnarray}
occupied by $\vq_1,\cdots,\vq_n$ that contribute to the estimation of 
angle-averaged $n$-th order polyspectrum.
Note that we can compute $V_{\tilde{n}}$ with the exact same method:
\be
V_{\tilde{n}}
=
\int\frac{d^3x}{(2\pi)^3} 
\prod_{j=1}^n
\iota_{k_j}(\vx)\,,
\label{eq:Vn}
\ee 
with
\be
\iota_{k_j}(\vx) = \int_{k_j} d^3q e^{-i\vq\cdot\vx}\,.
\ee
In this paper, we shall implement \refeq{Vn} to study the performance of the
parallel algorithm.

\subsection{On the dimensionality of $n$-th order polyspectra} 
\label{sec:dimNpcf}
At this point, a cautionary remark on the dimensionality of 
$n$-th order polygon is in order. 
While we only consider the `angle averaged' polyspectra in this 
paper, specifying the full shape of $n$-point ($n\ge3$) correlation function 
(or polyspectra, its Fourier counterpart) in the three-dimensional space would
require $3n-6$ real numbers. It is because adding one more point to 
a configuration of $(n-1)$ points introduces three additional real numbers 
(the coordinate of the new point relative to the $n-1$ points) to specify 
the configuration of $n$ points. Starting the recursion from bispectrum 
(or three-point correlation function) that requires three numbers, 
therefore, each configuration of $n$-point correlation function is specified 
by $3n-6$ real numbers. 

This argument does not apply for the two-point function ($n=2$) and 
the three point function ($n=3$) because of the underlying symmetry.
For the two-point correlation function, translational symmetry removes three
(we can always set the coordinate of the first point as origin) and the 
rotational symmetry removes two (correlation function must be independent of
the orientation of the second point) real dimensions. 
As a result, we are left with one real dimension for the power spectrum.
For the three-point correlation function, the rotational symmetry around the 
axis defined by existing two points removes one real dimension.
As a result, we are left with three real dimensions for the bispectrum.

Along the same reasoning, adding redshift-space distortion 
\citep{kaiser:1987} would require only two more real dimensions to the 
problem, because the coordinate of the newly added point is enough to 
determine the orientation of the point relative to the existing 
$n-1$ points. That is, from the two angles specifying the orientation of any three points relative to the line-of-sight direction, we can determine 
the relative orientation of all other points with respect to the line-of-sight
direction.

Although we shall not discuss further in this paper, one can also implement
the polyspectra estimator with the full dependence by using a similar 
method. For example, in addition to the amplitude of four wavevectors, 
the general trispectrum also depends on the diagonals 
$d_1 =|\mathbf{k}_1 - \mathbf{k}_2|$ and 
$d_2 = |\mathbf{k}_1 - \mathbf{k}_3|$, takes the form 
\ba
&T(k_1,k_2,k_3,k_4,d_1,d_2) 
\vs 
=&    \frac{V_f}{V_T(2\pi)^3}\int_{k_1}d^3q_1\cdots\int_{k_4}d^3q_4\int_{d_1}d^3p_1\int_{d_2}d^3p_2\delta_{D}(\mathbf{q}_{1234}) \nonumber
\\ &\times \delta_D(\mathbf{p}_1 - \mathbf{q}_1 + \mathbf{q}_2) \delta_D(\mathbf{p}_2 - \mathbf{q}_1 + \mathbf{q}_3) \delta_{\mathbf{q}_1}\delta_{\mathbf{q}_2}\delta_{\mathbf{q}_3}\delta_{\mathbf{q}_4} 
\vs
=&   
\frac{V_f}{V_T(2\pi)^3}
\int\frac{d^3x}{(2\pi)^3}
I_{k_4}(\vx)
\int\frac{d^3y}{(2\pi)^3}
\iota_{d_2}(\vy)
I_{k_3}(\vx+\vy)
\vs
&\times
\int\frac{d^3z}{(2\pi)^3}
\iota_{d_1}(\vz)
I_{k_1}(\vx-\vz-\vy)
I_{k_2}(\vx+\vz)\,.
\label{eq:Tkfull}
\ea
One may use, again, Fourier transformation to calculate the double convolution.
Note  that the result must reduce to the angle averaged trispectrum when
integrating over all possible $d_1$ and $d_2$.

\section{Parallelization} \label{sec:parallel}
\begin{figure}[ht]
\includegraphics[width=0.48\textwidth]{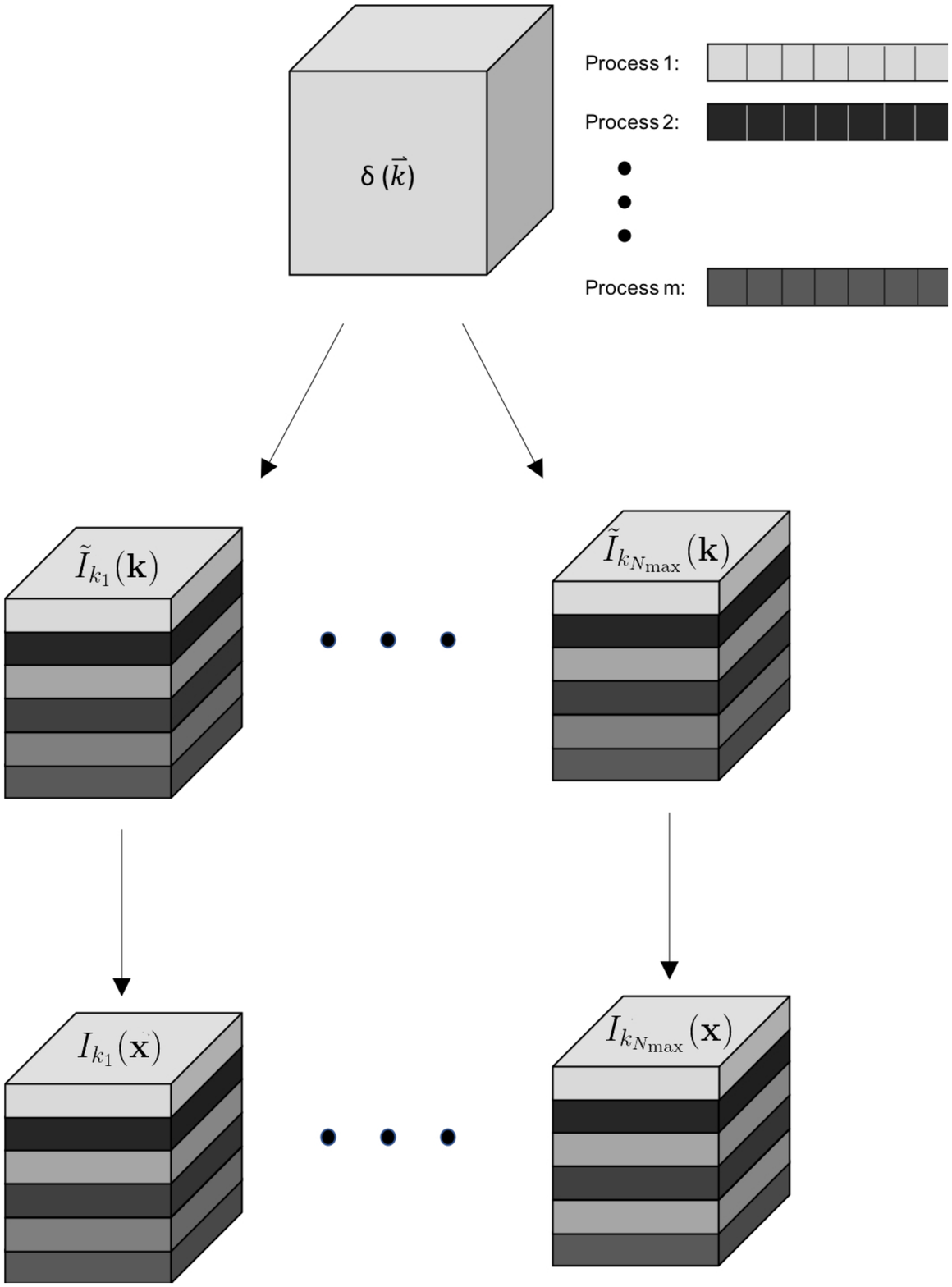}
\caption{
A visualization of the decomposition scheme. From the Fourier space density contrast field $\delta(\vk)$, we construct $N_{\rm max}$ Fourier data cubes $\tilde{I}_{k_i}(\vk)$ for $i=1$ to $N_{\rm max}$ and distribute the array to $m$ different processes depicted as different gray scale colors. The data cubes $\tilde{I}_{k_i}(\vk)$ are spread out over each process such that each one only has access to a single contiguous chunk of one index and the full range of the others. For example, the first process stores 
$\tilde{I}_{k_i}(k_F:k_{\rm 1max},k_F:k_{\rm max},k_F:k_{\rm max})$ from all $\tilde{I}_{k_1}$ to $\tilde{I}_{k_{N_{\rm max}}}$. In this distribution scheme, the integration in \refeq{estimator} is done mostly in each local process, and the inter-CPU communication is only required at the last moment when we compute the total sum. 
\label{fig:cube} 
}
\end{figure}
The key to the efficient parallelization of the polyspectra estimator is minimizing the inter-CPU communication, and we achieve that by using the slab decomposition scheme. 

The slab decomposition slices the data cube along a single dimension, generally the fastest varying index, and assigns each chunk to different process. As each process stores a full range of other dimensions and limited range of the sliced dimension, this separates the data into `slabs' of the domain hence the name slab decomposition. We show a visualization of this decomposition in \reffig{cube}. For the specific case of the polyspectra estimator, this decomposition takes the form of decomposing along the fastest varying Fourier grid dimension while the other two grid dimensions and the wavenumber dimension remain fully accessible in each process. For example, for the {\sf Julia} implementation, we assign each process $\tilde{I}_{k_i}(k_{1\rm min}:k_{1\rm max},k_{\rm F}:k_{\rm max}, k_{\rm F}:k_{\rm max})$, and for {\sf C} implementation, we assign each process $\tilde{I}_{k_i}(k_{\rm F}:k_{\rm max}, k_{\rm F}:k_{\rm max},k_{3\rm min}:k_{3\rm max})$ for all $i$ from 1 to $N_{\rm max}$.

The choice of slab decomposition makes reading data from a file very memory efficient as the main process can send data to the other processes partway through reading the data by simply sending each slab to the proper process as it is read in, or by having each process read in their own slab of data. This also means that at no point does the reading process need more than a single slab of data at once in memory which helps to keep the maximum memory usage low. This memory distribution is also natural for the Fourier transformation as FFTW \citep{fftw05} implements distributed Fourier transforms through slab decomposition and transposing, allowing for a very natural calculation without need to shuffle around memory between processes to get the right format. The only other step in the calculation is the product and sum, the product is always able to be done locally within each process and the sum is a trivially parallizable calculation with very little extra memory usage or communication as each processes calculates a local sum before the main processes collects and finalizes the sum.
All together this memory distribution scheme allows us to easily and quickly calculate this estimator with minimal memory footprint and very little inter-CPU communication. 

This method must be contrasted with the na\"ive parallelization that each process contains the full $\tilde{I}_{k_i}(\vk)$ and $I_{k_i}(\vx)$ for each wavenumber bin $k_i$. For this alternative case, however, the time saved in the Fourier transform step is dwarfed by the significantly more expensive final step where we calculate sums of products across wavenumbers. As a result, this somewhat more natural decomposition is significantly slower than the slab decomposition.

\section{Code Tests} \label{sec:calcnk}
In order to test the accuracy and performance of our 
implementation of the parallel estimator, we compute the number 
of polygons satisfying the homogeneity condition and 
the binning condition:
\ba
N_{n\rm -poly}
=& \frac{V_{\tilde{n}}}{V_f^{n-1}}
\vs
=& 
\frac{1}{k_F^{3n-3}}\int_{k_1}d^3q_1\cdots\int_{k_n}d^3q_n \delta_D(\mathbf{q}_{1\cdots n}),
\label{eq:Ncalc}
\ea
for which we have present an analytical estimation (see 
\refsec{polygon} for the detail).
Here, $V_f$ is the volume of the fundamental cell in the Fourier
space, which is related to the fundamental Fourier wavenumber
$k_F$ by $V_f = k_F^3$.

In practice, we estimate the number of polygons by setting 
$\delta(\vk)=1$ for all Fourier modes and estimating the
angle-averaged polyspectra. Therefore, counting the number of 
polygons requires exactly the same procedure as estimating 
the polyspectra.

\subsection{Number of Polygons}\label{sec:polygon}
To obtain the number of $n$-gons satisfying the binning conditions,
we need to evaluate the $3n$-dimensional integration in 
\refeq{Ncalc}. Again, using the Fourier representation of the 
Dirac-delta operator, we reduce the problem to the 
three-dimensional integration of multiplications of 
$\iota_{k}(\bfx)$ (\refeq{Vn}), which is
\be
\iota_{k} 
=
\int_k d^3q e^{i\bfq\cdot\bfx}
=
4\pi k^2 \left(\frac{\sin(kx)}{kx}\right) \delta k
+
{\mathcal O}(\delta k^3)\,.
\ee
Using the expression to leading order in the bin size $\delta k$, 
we further reduce the calculation of the $3n$-dimensional volume to 
the one-dimensional integration:
\ba
V_{\tilde n}
=&
\int \frac{d^3x}{(2\pi)^3}
\prod_{i=1}^n \iota_{k_i}(x)
\vs
=&
(4\pi \delta k)^n 
\int_0^\infty 
\frac{dx}{2\pi^2 x^{n-2}}
\prod_{i=1}^n
k_i \sin(k_ix)\,.
\ea
In \refapp{analytic}, We find a general expression for the radial
integration,
\ba
&\int_0^\infty \frac{dx}{x^{n-2}}\prod_{i=1}^n
\sin(k_i x)
=\frac{\pi}{2}
\left(
\prod_{i=3}^{n}
\int_{-k_i}^{k_i}\frac{d\kappa_i}{2}
\right)
\vs
&\times\left[
\delta^D\left(\sum_{i=3}^{n}\kappa_i+k_{1}-k_2\right)
-
\delta^D\left(\sum_{i=3}^{n}\kappa_i-k_{1}-k_2\right)
\right]\,,
\label{eq:Incalc}
\ea
and apply \refeq{Incalc} for $n=3$ and $n=4$ cases to find the 
number of triangles and quadrilaterals, which are given by
\be
N_{\rm tri} = \frac{1}{k_F^6}
\left\{
\begin{array}{cc}
4\pi^2 k_1k_2k_3\(\delta k\)^3 &,~k_1=k_2+k_3 \\
8\pi^2 k_1k_2k_3\(\delta k\)^3 &,~k_1<k_2+k_3
\end{array}
\right.\,,
\label{eq:Ntri}
\end{equation}
and 
\ba
N_{\rm quad}
=& \frac{1}{k_F^9} 8\pi^3
\left(\delta k)^4 k_1k_2k_3k_4(-k_1+k_2+k_3\right.
\vs
&\left.+3k_4 - |k_1-k_2-k_3+k_4|\right)\,.
\label{eq:Nquad}
\ea

\subsection{Result}\label{sec:result}
\begin{figure*}[ht]
\begin{center}
\plotthree{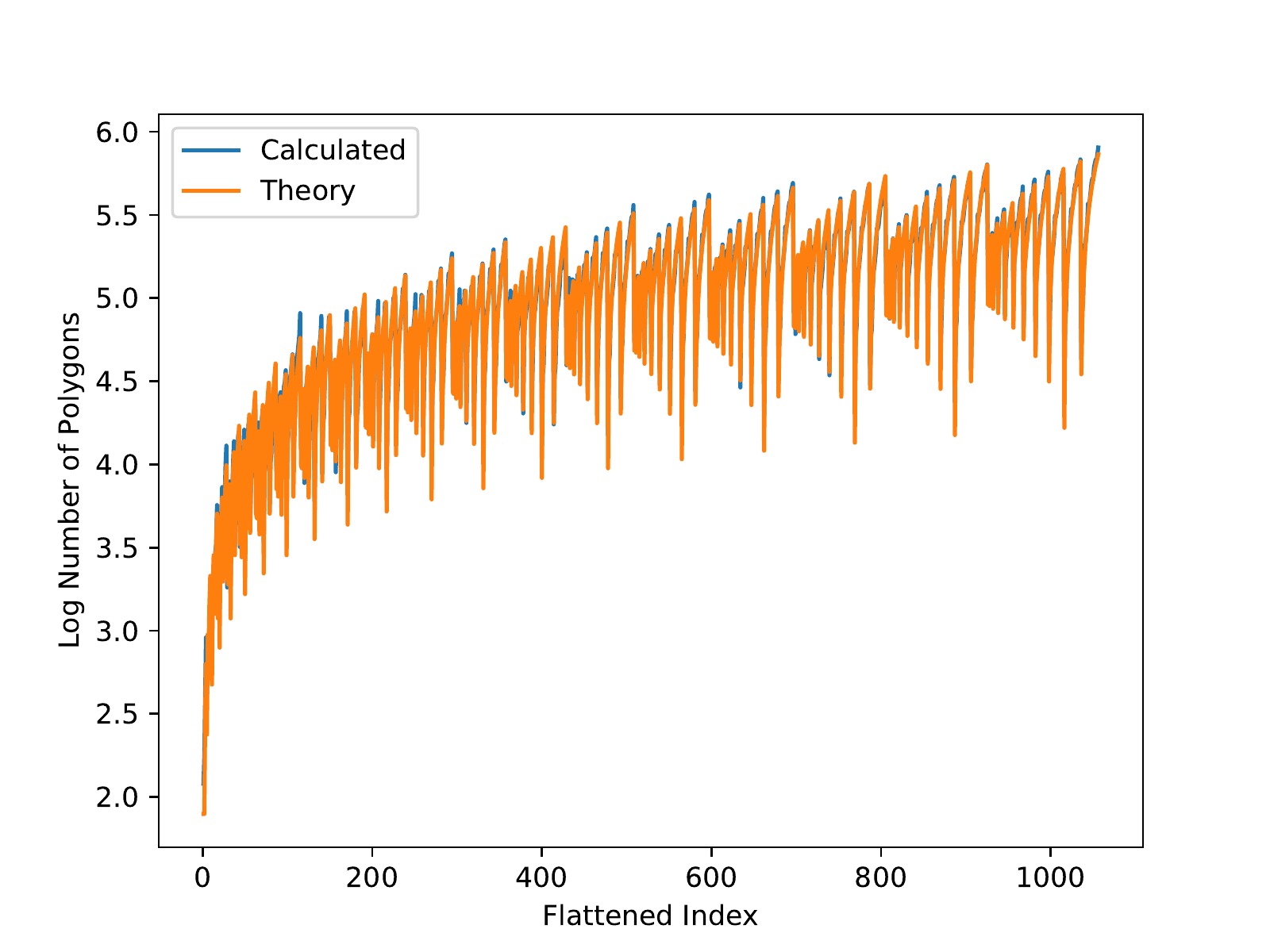}{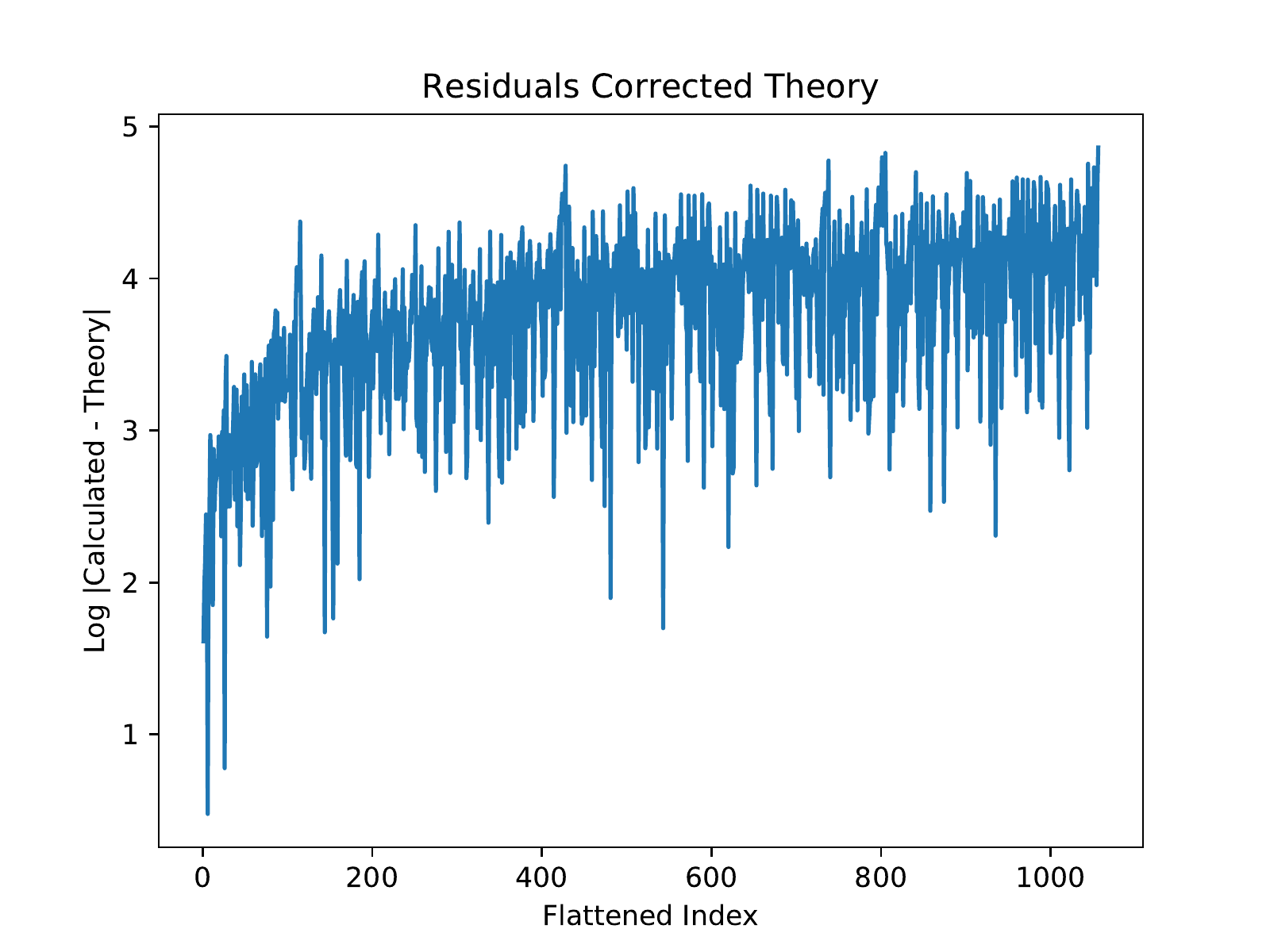}{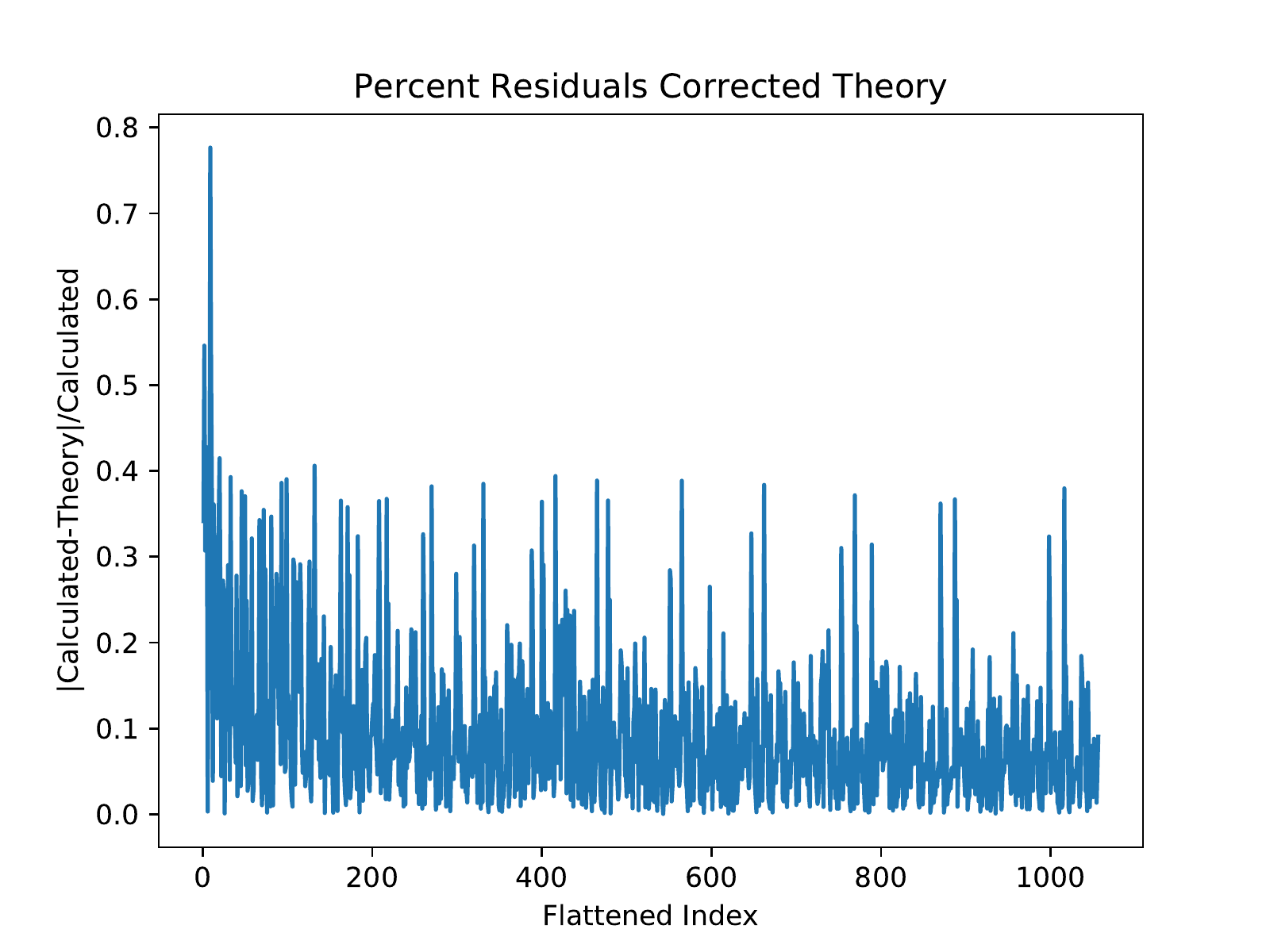}
\end{center}
\caption{\label{fig:flat1} {\it Left}: The number of triangles satisfying the binning conditions measured from the parallel bispectrum estimator. The blue curve is the output from the numerical estimation, and the orange curve is according to our analytical estimation in \refeq{Ntri} (see \refapp{analytic} for the derivation). {\it Middle}: The difference between the numerical measurement and theoretical estimation. {\it Right:} The fractional difference between the two. Besides the largest-scales (earlier indices) that we expect the analytical estimation breaks down, the analytical estimation in \refeq{Ntri} is good to about $20~\%$ accuracy.}
\end{figure*}
\begin{figure*}[ht]
\begin{center}
\plotthree{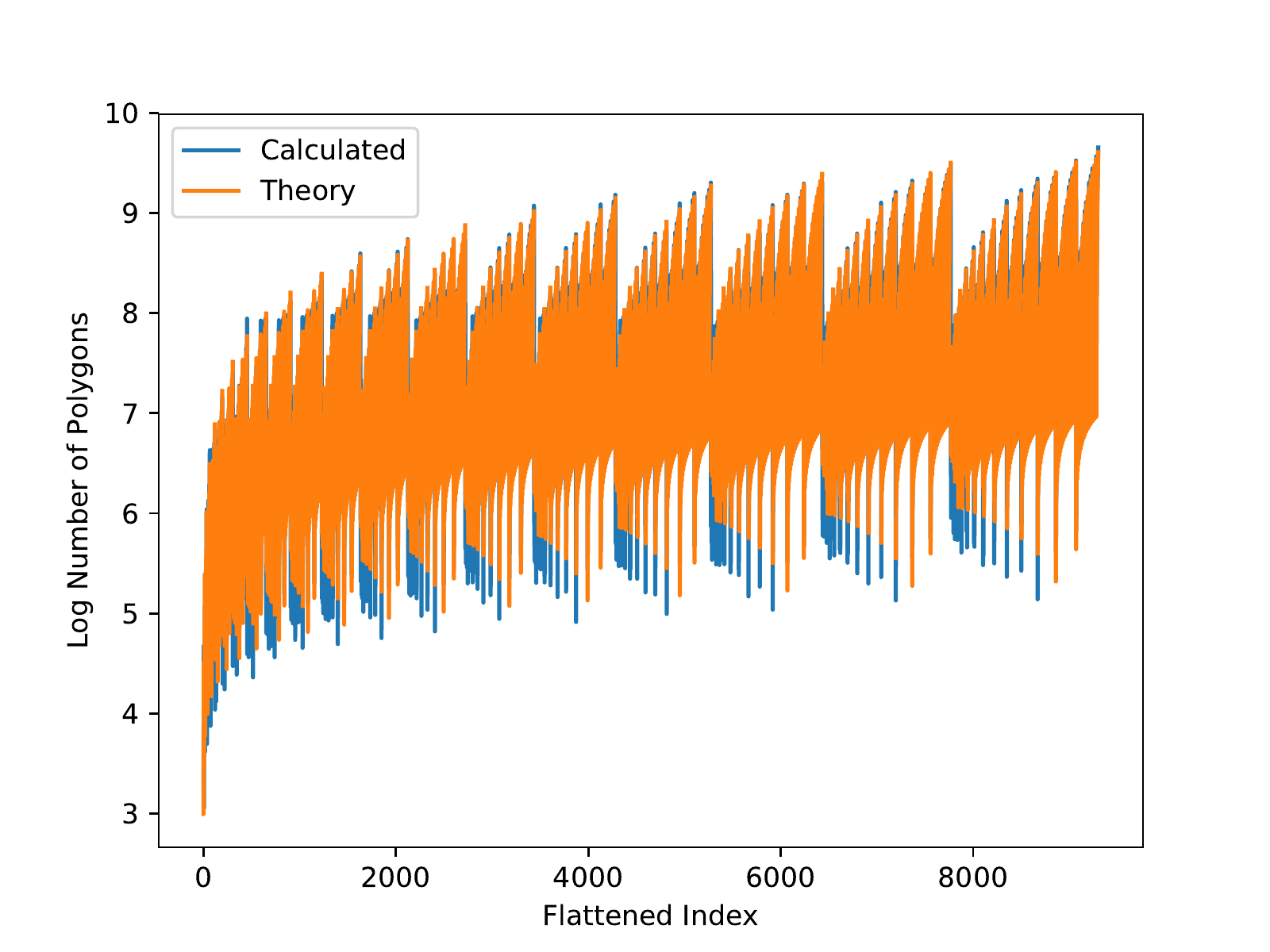}{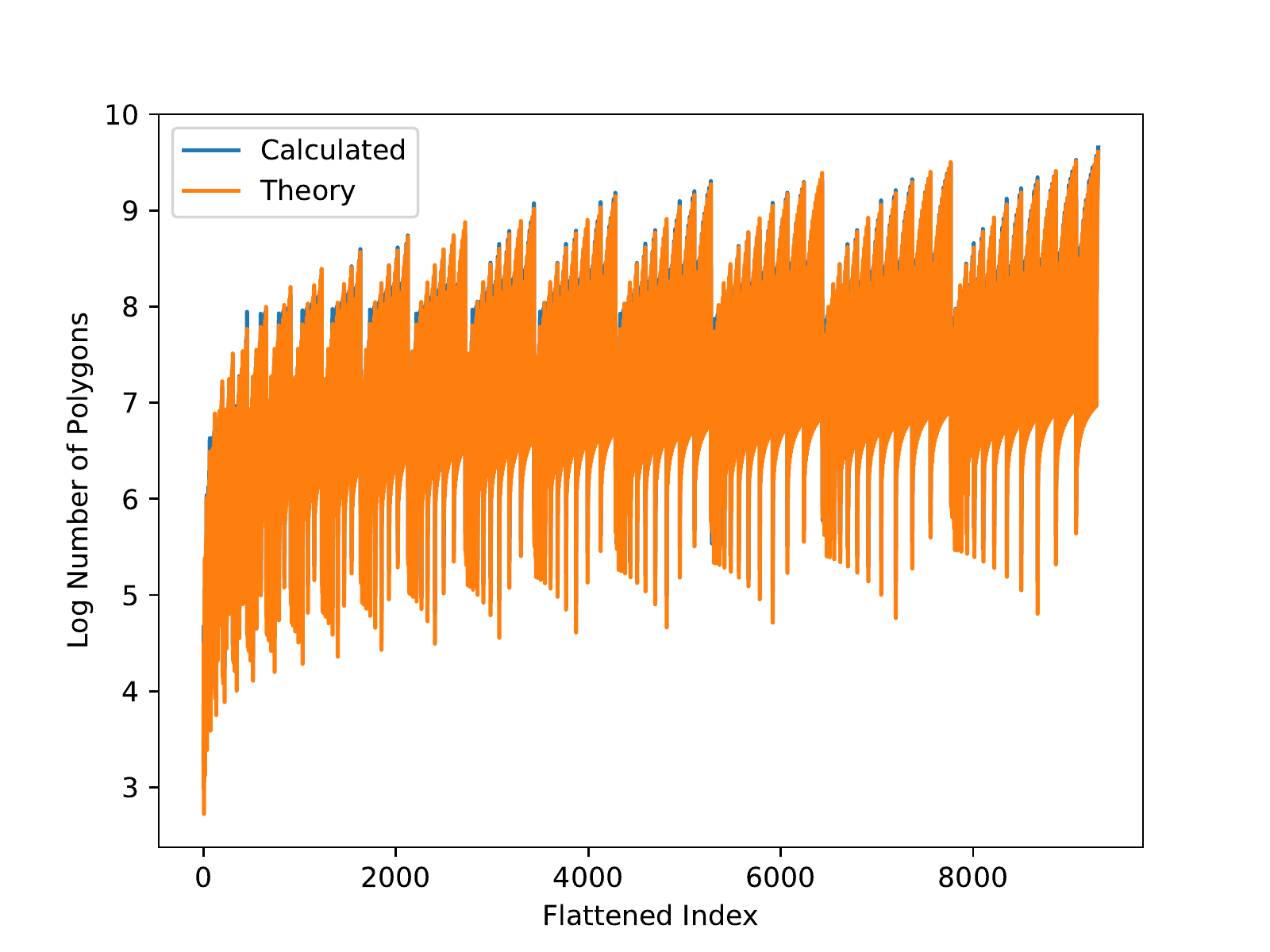}{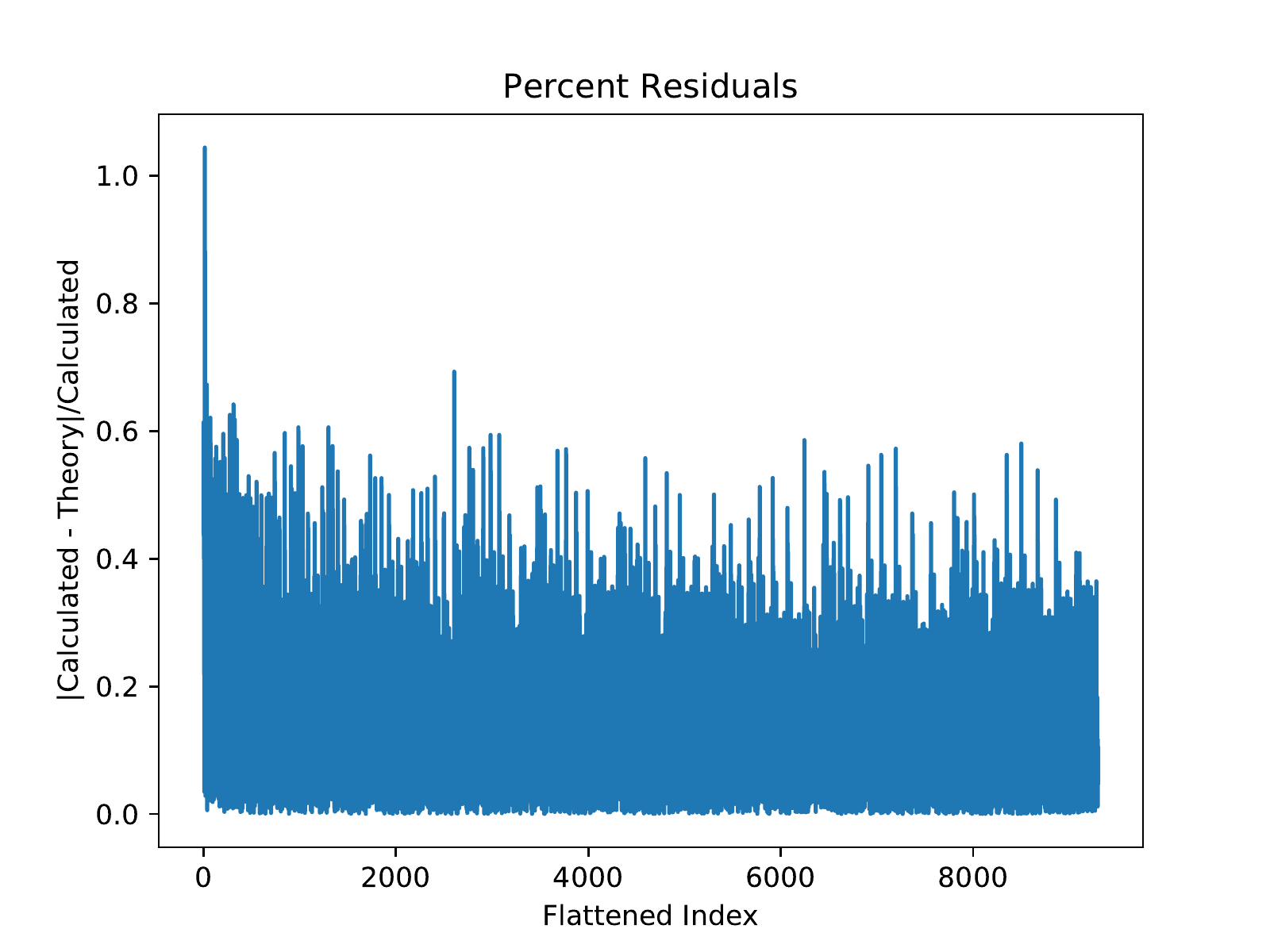}
\end{center}
\caption{\label{fig:triflat}
{\it Left}: The number of quadrilaterals satisfying the binning conditions measured from the parallel trispectrum estimator. The blue curve is the output from the numerical estimation, and the orange curve is according to our analytical estimation in \refeq{Nquad}. {\it Middle}: The same numerical result (blue) along with the improved analytical calculation, using exact expression for $\iota_k(x)$ (see, \refeq{iota_exact}) and calculate the correction for the co-linear quadrilaterals as in \refeq{tricol}. {\it Right}: The fractional difference between the two curves in the middle panel. 
The small-scale (large wavenumber) discrepancies are consistently about $40~\%$.}
\end{figure*}

In this section, we shall present number of polygons that 
we calculated using the parallelized polyspectra estimator and 
compare the result with the analytical estimation that we find
in \refsec{polygon}. We shall show the result for 
triangles ($n=3$), quadrilaterals ($n=4$), pentagons ($n=5$),
and hexagons ($n=6$) that are estimated using, respectively, the 
parallel estimators of bispectrum, trispectrum, quadspectrum, and 
pentaspectrum.

To show the full shape dependence of the number of polygons as a 
function of their side length (wavenumbers), we need to visualize
them in $n$-dimensional space of $(k_1,k_2,\cdots,k_n)$. For our
purpose, however, to compare the outcome of the parallel 
computation and analytical estimation, it suffices to flatten 
$n$-dimensional data points to the one-dimensional ordered 
sequence. Following the convention in \cite{2016MNRAS.455.2945M},
we impose the condition $k_i \geq k_{i+1}$ to avoid
duplication and form the sequence by taking the row-major 
order of the $n$-dimensional points ($k_n$ is the fastest varying
dimension).

In all comparison plots that we present here, the $x$ axes show the
index of the $n$-dimensional point $(k_1,\cdots,k_n)$ in the
flattened sequence, and $y$ axes show the number of polygons whose
$n$ sides are $(k_1,\cdots,k_n)$. We use blue color to show the 
numerical result and orange color to show the analytical 
estimation.

First, we show the number of triangles using the parallel 
bispectrum estimator in \reffig{flat1}. The left panel shows the 
numerical result (blue) and analytical prediction (\refeq{Ntri}),
while the middle and right panels show the absolute number 
(middle) and fractional (right) differences. On large scales (or small
indices), we expect the analytical estimation is not accurate because 
the bin's size is of order the wavenumber ($k_i\simeq \delta k$). 
For small scales, although the absolute residuals increases with 
wavenumbers (the lowest order error term goes as $\prod k_i$ just like
the $N_{\rm tri}$ in \refeq{Ntri}), the fractional residual between 
the numerical and analytical estimation are consistently off by 
$10\sim20\%$, and the biggest differences happen for the folded triangles
($k_1=k_2+k_3$). This result suggests that the variance of bispectrum
estimated by using \refeq{Ntri} (for example, in the Fisher 
information forecast in \citet{pbsreview}) must be off by the 
same factor. For an accurate estimation of the variance of the 
bispectrum, one must use the numerically estimated number of triangles.

\begin{figure*}[ht]
\plotthree{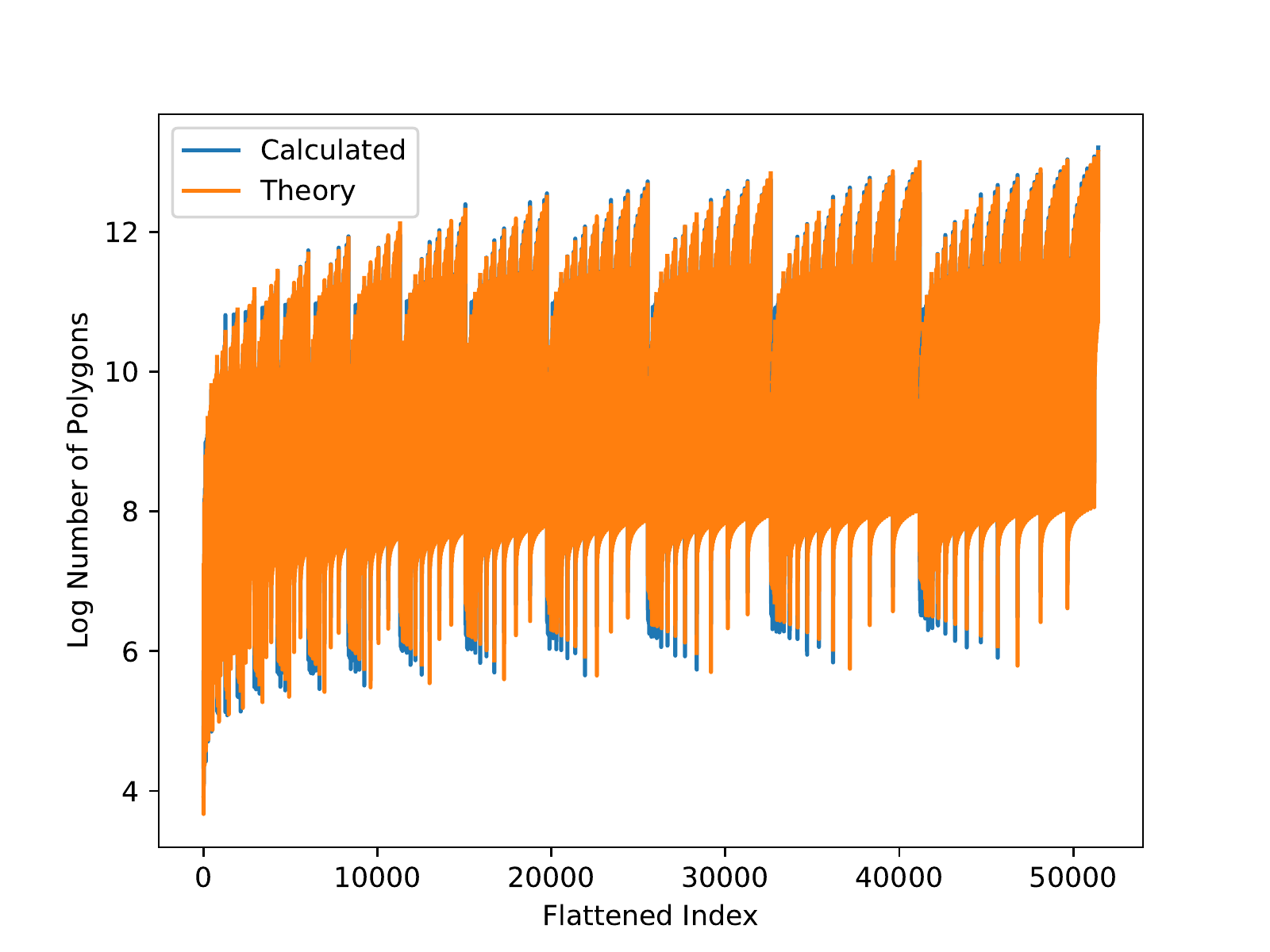}{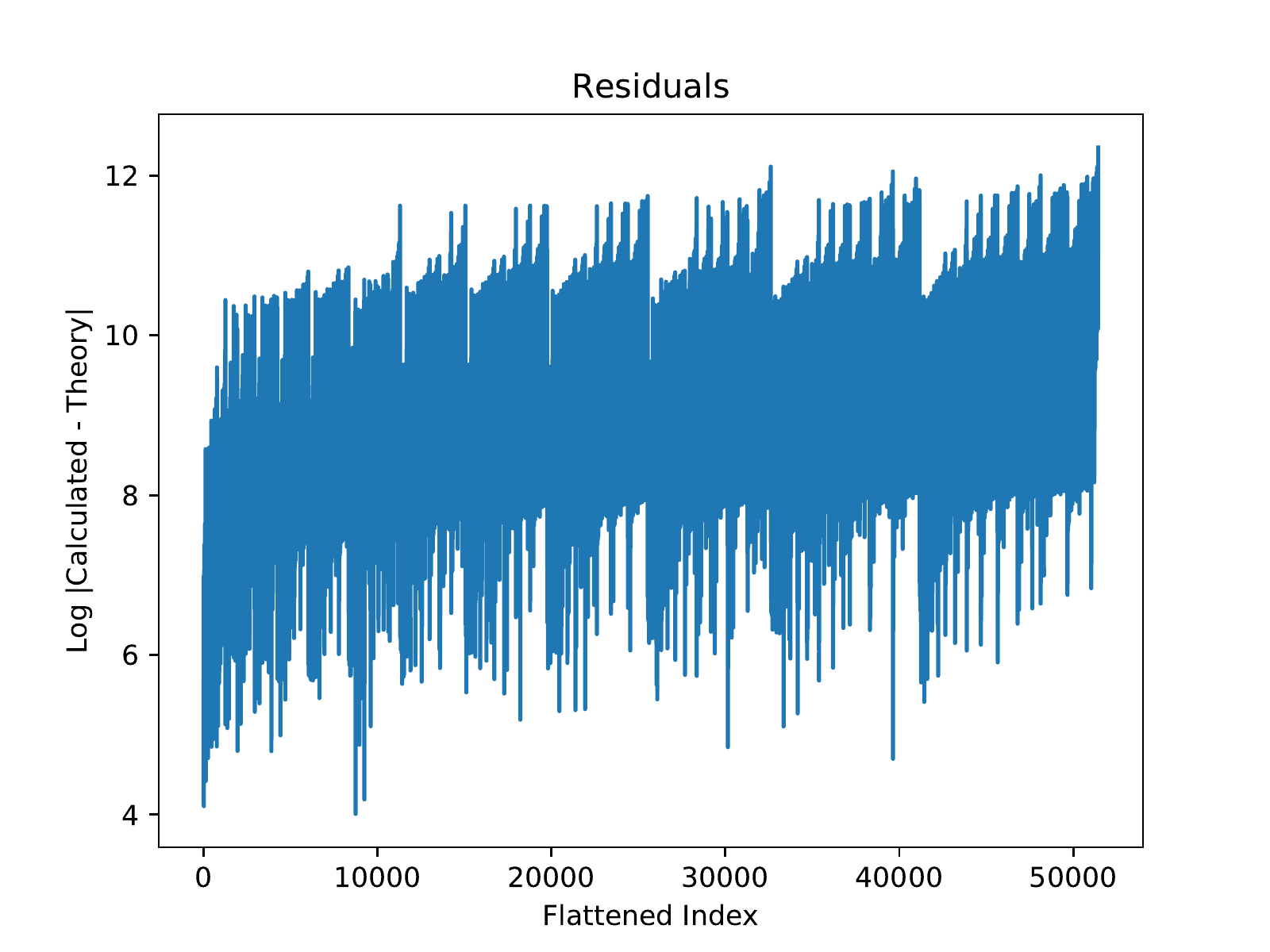}{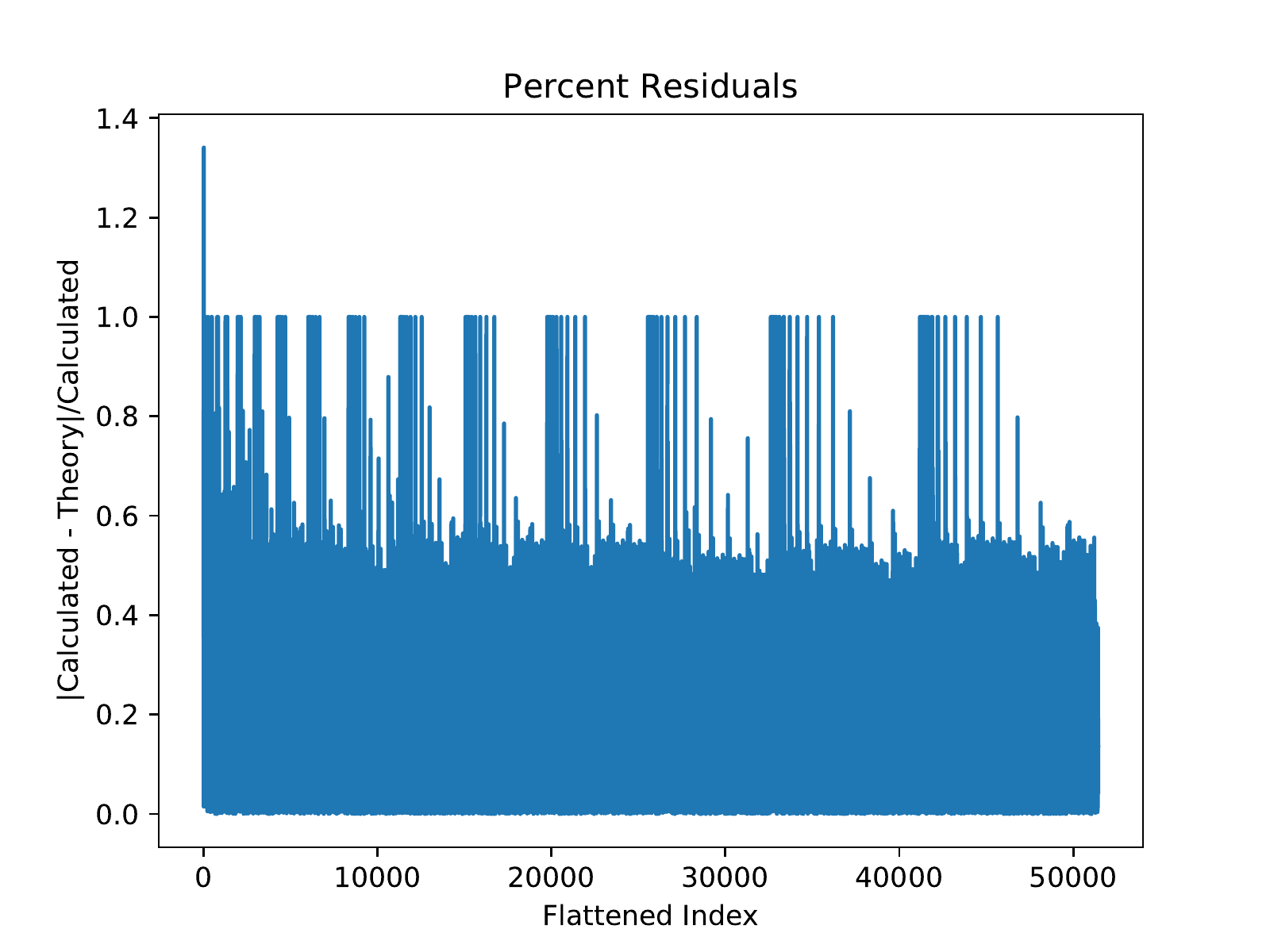}
\caption{\label{fig:quadflat} The flattened number of pentagons/Angle Averaged Quadspectrum of unity. The orange is according to the number predicted by \refapp{analytic} and the blue curve is the output of the estimator run on a grid of unity on a spherical shell. The Quadspectrum shows very similar error behaviour to the Trispectrum and for much the same reasons cannot be practically used when calculating actual Quadspectrum.}
\end{figure*}
\begin{figure*}[ht]
\plotthree{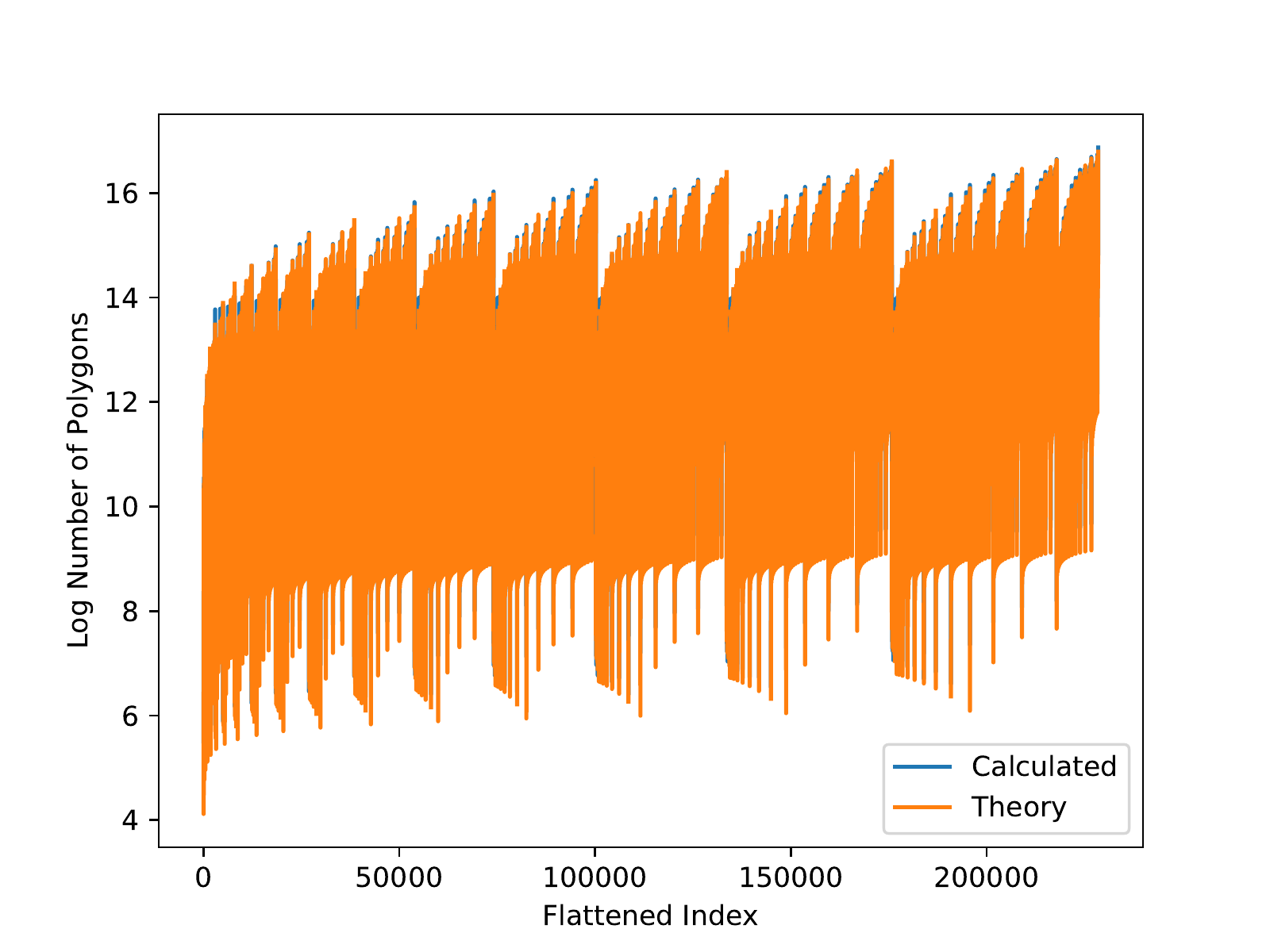}{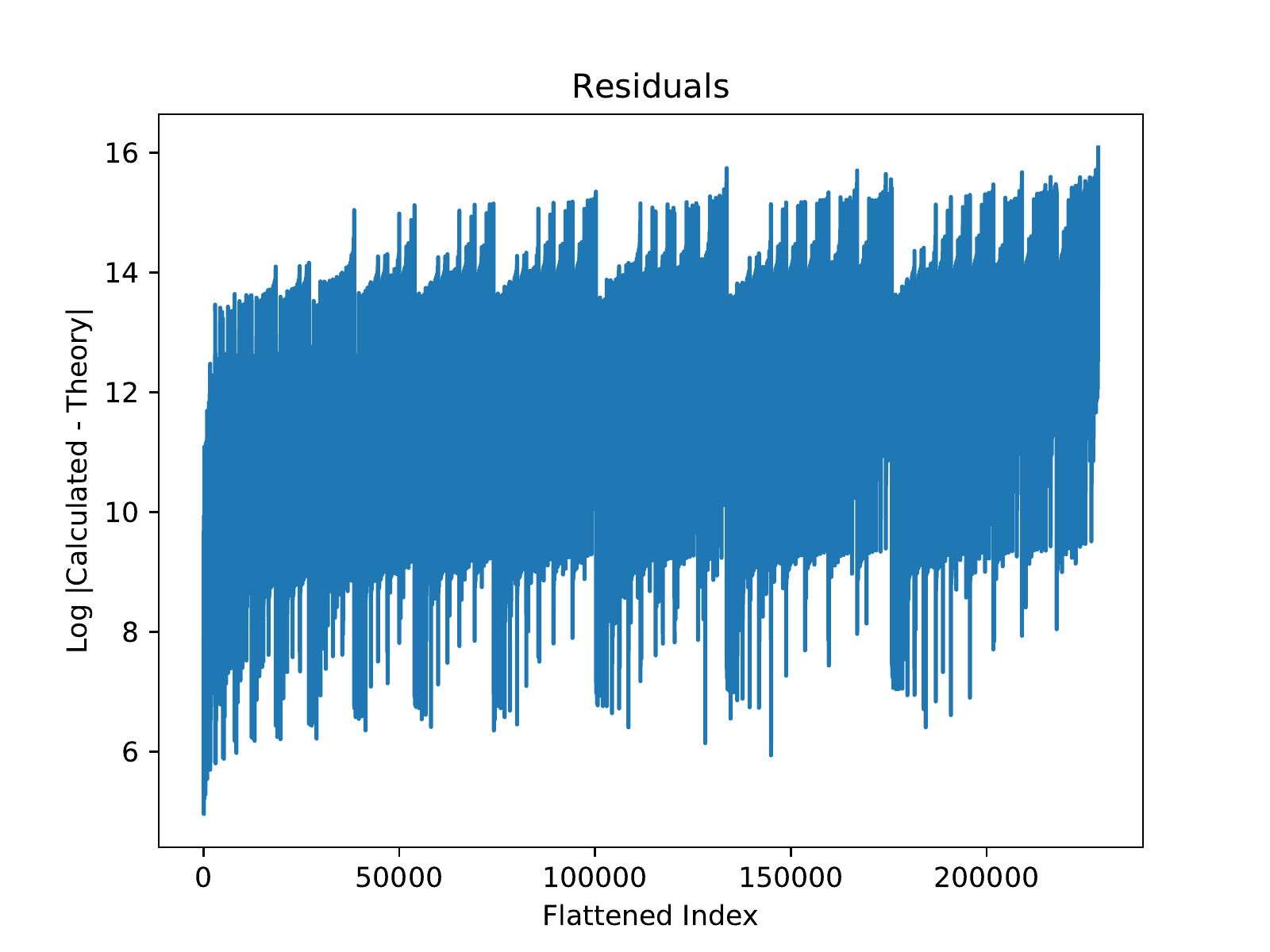}{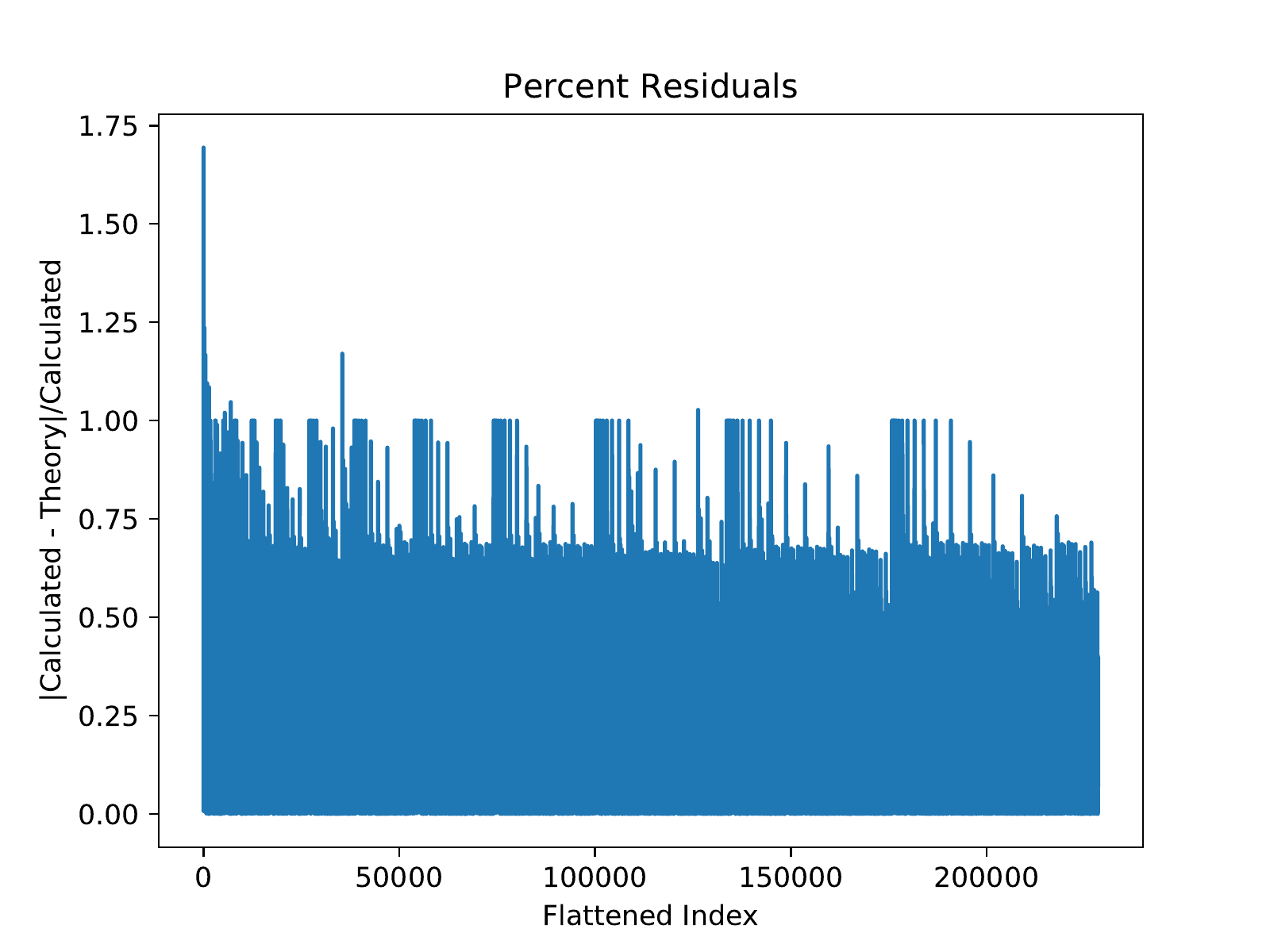}
\caption{\label{fig:pentflat} The flattened number of hexagons/Angle Averaged Pentaspectrum of unity. The orange is according to the number predicted by \refapp{analytic} and the blue curve is the output of the estimator run on a grid of unity on a spherical shell. The Pentaspectrum displays near identical error behaviour to the other high order polyspectra.}
\end{figure*}
We perform the same analysis for the number of quadraliterals by using
the parallel estimator for trispectrum. The left panel of \reffig{triflat}
shows the comparison between the numerical result and the analytical 
estimation in \refeq{Nquad}. It turns out that biggest the 
discrepancies are for the co-linear quadrilaterals ($k_1=k_2+k_3+k_4$)
where the analytical formula, \refeq{Nquad}, predicts zero. To remedy the 
situation, we add the exact 
corrections for the co-linear quadrilaterals, \refeq{tricol}, and 
the analytical estimate indeed matches the numerical calculation better,
as shown in the middle panel of \reffig{triflat}.

We can easily extend the parallel estimator for the bispectrum and 
trispectrum to measure any $n>4$ polyspectra. We extend the estimator to
$n=5$ and $n=6$ order polyspectra, and show the result in
\reffig{quadflat} and \reffig{pentflat}, respectively. In all cases 
the fractional errors between numerical calculation and analytical 
estimate are dominated by a collection of 100\% error terms. 
These occur, just as the co-linear case for the quadrilaterals, 
because the leading-order approximation in the analytical calculation
yields zeros when the actual calculation yields small but 
not-zero numbers. This happens for all $n>4$ cases, and it further
reinforces the need for manually calculating the number of polygons when 
estimating the polyspectra, or estimating their variance.

\subsection{Code performance} \label{sec:nkperformance}
In this section, we present the performance of the parallel algorithm.
All of our benchmarking work was done with 2.2 GHz Intel Xeon Processors
on nodes with 40 CPU/node, 1 TB of RAM, 10 Gbps Ethernet and FDR 
Infiniband.

\begin{figure}
\plotone{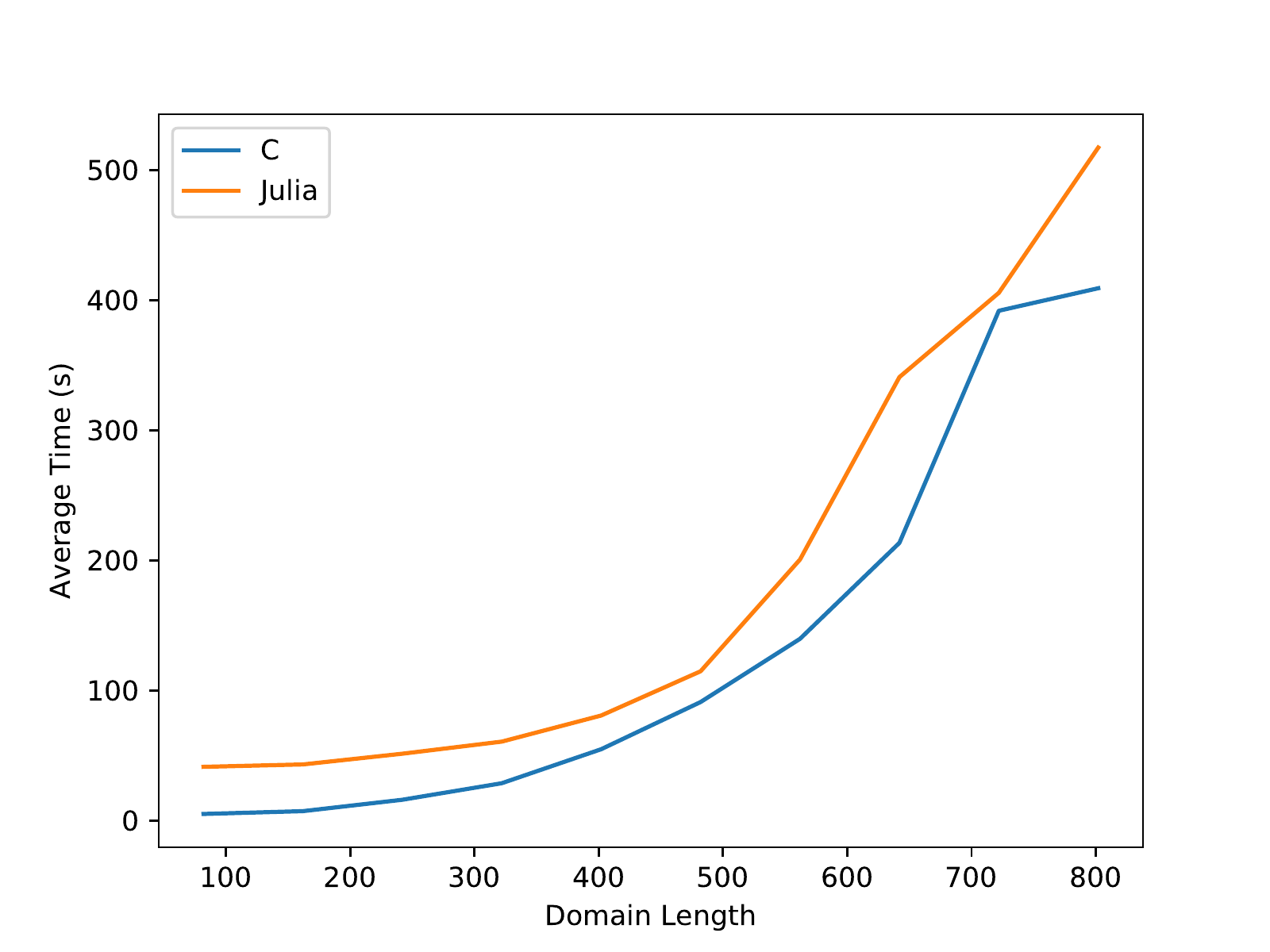}
\caption{\label{fig:bench} The total time spent calculating the Bispectrum versus the length of one grid dimension for 40 processors. It roughly follows the $\mathcal{O}(N^3)$ you would expected given the sum in the algorithm with fixed $N_{\rm max}$, modulo the overhead dominated small grid sizes. Each grid size was tested 50 times and then averaged. While the Julia version is slightly slower than the C version, the time difference is relatively consistent with around a constant 30 second offset for small grid sizes and is mostly negligible for large grid sizes given the overall length of the computation.}
\end{figure}

We have implemented two versions of parallel bispectrum estimator, one 
using {\sf C} and the other using {\sf Julia}. These two pieces of software 
perform the same algorithm. In \reffig{bench}, we compare the performance
of the two implementations. Overall, as a function of domain length, 
the size of one-dimensional Fourier grid $N$ and the maximum 
wavenumber $N_{\rm max}=k_{\rm max}/k_F$, the performance scales as
$\mathcal{O}(N^{3}N_{\rm max}^{n})$, as we have analyzed in 
\refsec{serial}. If we apply the aliasing condition which requires that 
$N > nN_{\rm max}$ we can rewrite the scaling as 
$\mathcal{O}(n^3N_{\rm max}^{n+3})$. 

The {\sf C} implementation (blue) out performs the {\sf Julia} 
implementation (orange) for all domain lengths. The improvement, however, 
stays constant at about 30 seconds offset for reasonable grid sizes. As it is easier to read and
modify the {\sf Julia} version, we have extended the software to higher 
order polyspectra estimator only using {\sf Julia}. 

\begin{figure*}
\plottwo{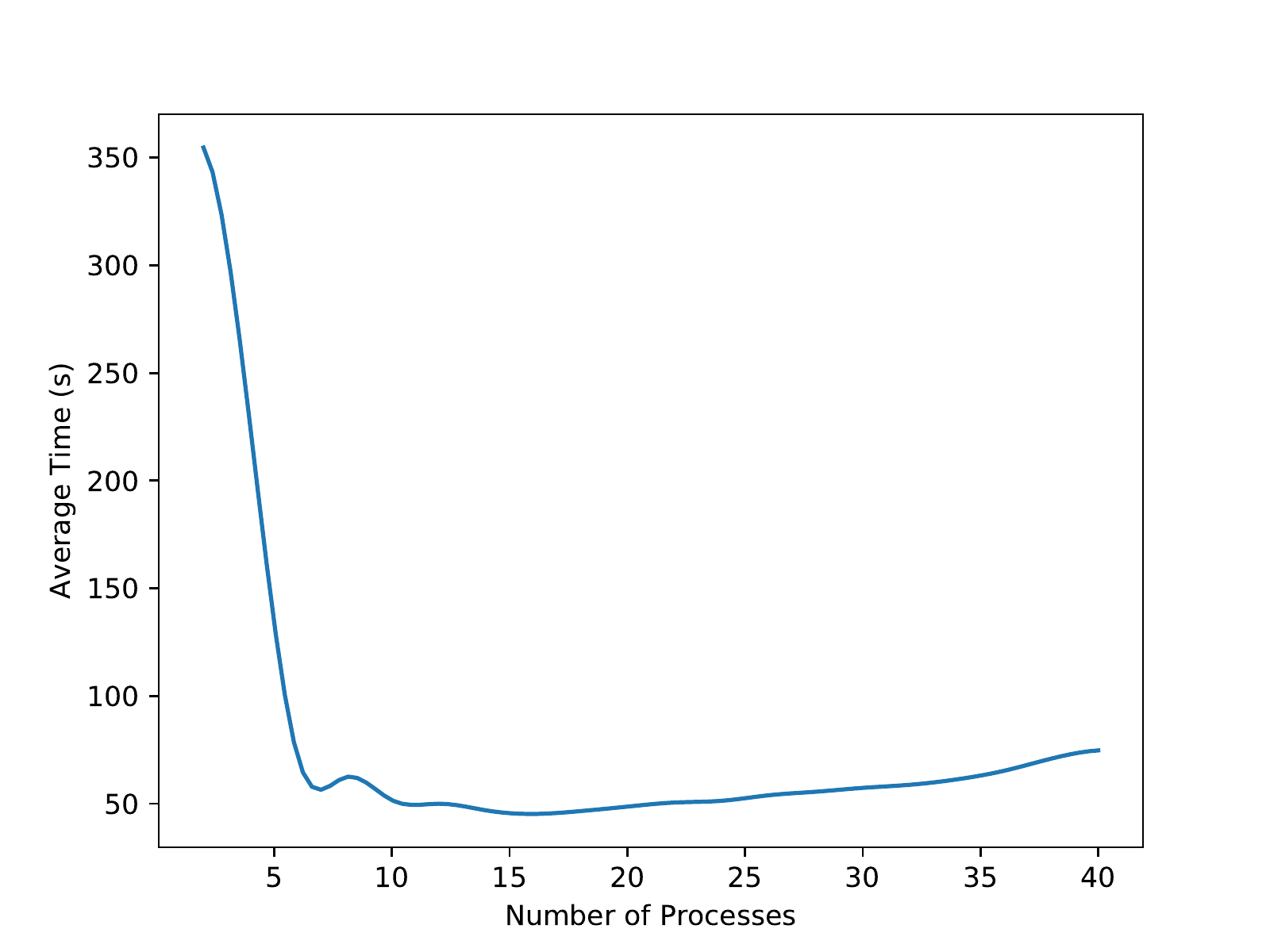}{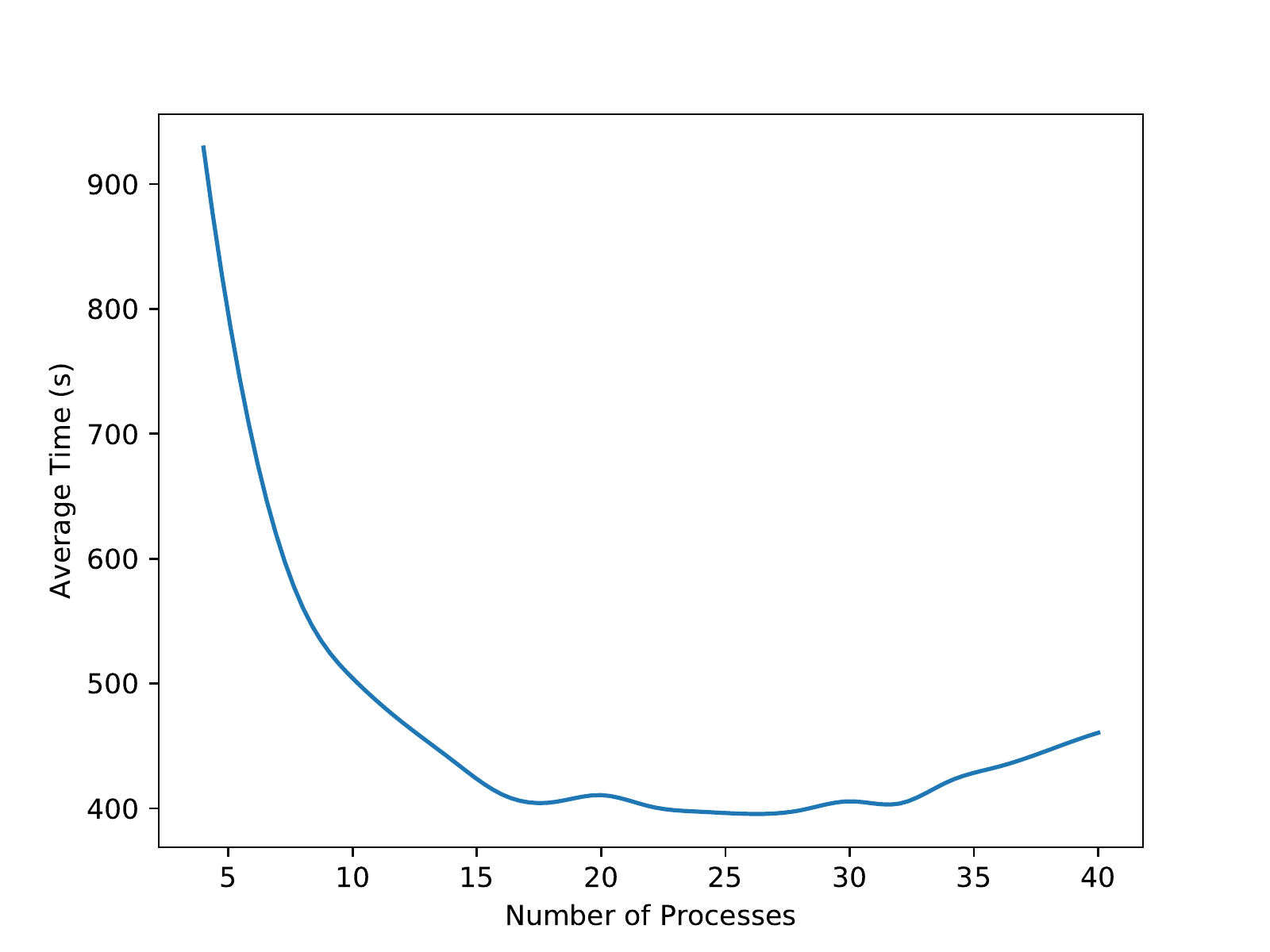}
\caption{\label{fig:cpu} Left: The scaling of an N=256 length grid over various numbers of processes. There is clear gains when initially increasing the number of processes but at a certain point overhead begins to dominate causing a slight slowdown. Right: The scaling of an N=512 length grid over various numbers of processes. It clearly takes a much larger number of processes to reach maximum efficiency, and is fast for larger number of processes. This scaling continues as grid size increases, meaning for reasonable grid sizes and number of processes it should be maximally efficient. Each number of processes was tested 50 times and then averaged.}

\end{figure*}

\reffig{cpu} shows the performance of our {\sf Julia} implementation
as a function of number of processors. Here, we use the domain size of
$N=256$ (left panel) and $N=512$ (right panel). Initially, adding more 
CPUs improves the performance and reduces the execution time, but the
improvement stops at some point. In fact, when the grid size per CPU is
sufficiently small then adding more processes actually slows down the 
performance; the increased overhead time cancels the small gains in 
extra discretization. 

How long does it take to estimate the bispectrum? In order to avoid the
aliasing effect, the grid size must satisfy $N > 3N_{\rm max}$ for
estimating the bispectrum. Therefore, probing the bispectrum up to
$N_{\rm max}\simeq 340$ requires the one-dimensional Fourier grid of
$N=1024$; we can estimate the bispectrum in 20 minutes using 40 cores. 

For a more concrete example, the HETDEX survey 
\citep{HETDEX:2008} has $k_F \approx 0.0045$ $h$/Mpc. Measuring
the galaxy bispectrum to $k_{\rm max} = 0.2$ $h$/Mpc only
requires $N_{max}=45$. With the parallel version of Scoccimarro
estimator, we can measure the galaxy bispectrum for HETDEX in 
$\sim$40 seconds with 2 GB of memory distributed over 40 
processors. 
Here, we did not run the in-place FFTW, and kept both $I_{k_i}(\bfx)$ and $\tilde{I}_{k_i}(\bfq)$.
Similarly, WFIRST \citep{WFIRST:2015} has $k_F \approx 0.0028$ $h$/Mpc corresponding to
$N_{\rm max}=75$ for $k_{\rm max} = 0.2$ $h$/Mpc. 
For our implementation over 40 processors this corresponds to $\sim$50 seconds and 11 GB of memory distributed over 40 processors to calculate the galaxy bispectrum for WFIRST.

With the scaling that we have discussed in \refsec{serial}, we can extend
the previous example to find a general relationship. For a cosmological
survey with volume $V$ and maximum wavenumber $k_{\rm max}$, we 
calculate the number of one-dimensional sampling point as
\be
N_{\rm max} = \frac{k_{\rm max}}{k_F} = \frac{1}{2\pi}
k_{\rm max}V^{1/3}\,.
\ee
Because the performance of the Scoccimarro estimator scales as
$\mathcal{O}(N_{\rm max}^6)$, the estimated computing time becomes
\be
t_{Bispec} \approx 40 \(\frac{k_{\rm max}}{0.2 h/\text{Mpc}} \)^6 \(\frac{V}{2.7 \text{ Gpc}^3/h^3} \)^2 \text{sec},
\ee
over 40 processors. For the higher-order polyspectra, we can extend
this relationship by 
\be
t_{Pn} \sim \frac{n^3}{27} \(\frac{k_{\rm max}}{0.2 h/\text{Mpc}} \)^{n+3} \(\frac{V}{2.7 \text{ Gpc}^3/h^3} \)^{(n+3)/3} \text{min},
\ee
again, over the 40 processors. This relationship matches
with the execution time for the polyspectra that we have measured for
generating the figures in this paper.

\section{Conclusion} \label{sec:conclusion}

In this paper, we present a novel parallel algorithm for estimating 
polyspectra of a complex field given on three-dimensional regular grids.
While inheriting the efficiency of the Scoccimarro estimator, the parallel
algorithm alleviates its rather stringent memory requirement by distributing 
the array into multiple processors. In addition, the key for efficient 
parallization is the slab-decomposition scheme that maintains only low level 
of inter-CPU data communication, avoiding massive data transportation.

Although we have only presented the case for the monopole (angle-averaged)
polyspectra as a function of their side lengths (that is, the 
wavenumbers $k_i$), for estimating the redshift-space galaxy bispectrum, 
one can easily extend the method to the angular multipoles by applying the slab decomposition to the multipole-weighed fields $\delta_n(\bfq)$ as described 
in \citet{scoccimarro:2015}.
Beyond the galaxy bispectrum, measuring the polyspectra with the full 
(3$n$ - 6)-parameter dependencies requires little more elaboration, as the 
calculation involves convolution instead of matrix inner product (\refeq{Tkfull}). We shall leave this as a future project.

With the {\sf Julia} implementation, we have demonstrated that for the 
galaxy surveys such as HETDEX and WFIRST,
we can estimate the galaxy bispectrum down to $k\simeq 0.2\,{h/{\rm Mpc}}$ in about a minute, and even higher-order polyspectra in less than an hour time scales. With these execution time scales, it is feasible to study the high-order
polyspectra by directly measuring them from the massive mock galaxy catalogs \citep{Monaco/etal:2013,Kitaura/etal:2014,Avila/etal:2015,Izard/etal:2016,Munari/etal:2017,Agrawal/etal:2017,Taruya/etal:2018}. If it is not for their own sake, studying the galaxy trispectrum 
(the four-point correlation function in Fourier space) and the galaxy 
pentaspectrum (the six-point correlation function) shall certainly elucidate
the study of covariance matrix for, respectively, the galaxy power spectrum
and the galaxy bispectrum.

We have stored our {\sf Julia} implementation for the $n$-point 
polyspectra at (\url{https://github.com/JosephTomlinson/PolyspectrumEstimator}), 
and those who are interested in {\sf C} implementation for the bispectrum 
estimator may contact the authors.

\acknowledgments

We thank Emiliano Sefusatti for useful discussion. DJ and JT were supported at Pennsylvania State University by NSF grant (AST-1517363) and NASA ATP program (80NSSC18K1103).

\bibliographystyle{aasjournal}
\bibliography{main}

\appendix
\section{Polyspectrum Estimator in Discrete Fourier Transformation} \label{app:sum}
In the main manuscript, we have presented the polyspectra estimator in an
contiguous limit, as an integral form. In the practical implementation,
we have converted the estimator into a discrete summation as we present here.
For our discrete implementation, following the {\sf FFTW} convention, we use
the unnormalized Discrete Fourier Transform and explicitly pull out all 
normalization factors. 
For more details, authors refer to Chapter 7 of \cite{jeong:2010}.
Here, we note the discrete version of variables by square bracket $[\bfx]$.

Let us first summarize the continuous version of the polyspectra  estimator.
Starting with the integral form of the general estimator \refeq{estimator}
\begin{equation}
\tilde{P}_n(k_1\cdots,k_n) = \frac{V_f}{V_{\tilde{n}}(2\pi)^3} \int_{k_1}d^3q_1\cdots\int_{k_n}d^3q_n \delta_D(\mathbf{q}_{1\cdots n})\prod \delta(\mathbf{q}_i)\,.
\end{equation}
We convert the Dirac-delta operator into its Fourier representation form:
\begin{equation} \label{eq:dirac}
\delta_D(x) = \int\limits_{-\infty}^\infty \frac{d^3k}{(2\pi)^3}e^{ik\cdot x}\,,
\end{equation}
which, when input back into the estimator, results in
\begin{equation}
\tilde{P}_n(k_1\cdots,k_n) = \frac{V_f}{V_{\tilde{n}}(2\pi)^3} \int \frac{d^3x}{(2\pi)^3} \prod I_{k_i}(x),
\end{equation}
where
\begin{equation}
I_{k_i}(x) 
= \int_{k_i}d^3q_i e^{i\bfx\cdot \bfq_i}\delta(\bfq_i)
= \int d^3q_i e^{i\bfx\cdot \bfq_i}\tilde{I}_{k_i}(\bfq_i)\,,
\end{equation}
is the Fourier Transform of $I_{k_i}(\bfq)$ which is defined to be the same
as $\delta(\bfq)$ within a spherical shell with Fourier-space radius 
$|\bfq|\simeq k_i$, and zero otherwise.

In the {\sf FFTW} convention of un-normalized DFT, we find
\be
I_{k_i}[\bfx] =  
\frac{(2\pi)^3}{N^3}
{\rm FFTW}
\left(
    \tilde{I}_{k_i}[\bfq]
\right)\,,
\ee
with which we get a partially discretized estimator
\begin{equation}
    \tilde{P}_n(k_1\cdots,k_n) = \frac{V_f}{V_{\tilde{n}}(2\pi)^3}\left(\frac{(2\pi)^3}{N^{3}}\right)^n \int \frac{d^3x}{(2\pi)^3} 
    \prod_{i=1}^n 
{\rm FFTW}
\left(
    \tilde{I}_{k_i}[\bfq]
\right)\,.
\end{equation}
We then fully discretize the estimator by converting the final integral 
to a sum:
\ba
\tilde{P}_n(k_1\cdots,k_n) 
=& 
\left(
\frac{V_f^{n-1}}{V_{\tilde{n}}}
\right)
\frac{(2\pi)^{3n}}{V_f^{n-2} N^{3n}}
\frac{V}{(2\pi)^6N^3}
\sum\limits_{[\bfx]}
\prod_{i=1}^n 
{\rm FFTW}
\left(
    \tilde{I}_{k_i}[\bfq]
\right)
\vs
=&
\left(
\frac{V_f^{n-1}}{V_{\tilde{n}}}
\right)
\frac{V^{n-1}}{N^{3n}} \frac{1}{N^3}
\sum\limits_{[\bfx]}
\prod_{i=1}^n 
{\rm FFTW}
\left(
    \tilde{I}_{k_i}[\bfq]
\right)\,.
\ea
Finally, the quantity in parenthesis $\frac{V_f^{n-1}}{V_{\tilde{n}}}$ is 
the reciprocal of the number of polygons, and we can compute them by applying the same polyspectra estimator to a Fourier grid of unity. Note, however, 
that the normalization is different here.
\begin{equation}
\frac{V_{\tilde{n}}}{V_f^{n-1}} 
=
\frac{1}{V_f^{n-1}} \int_{k_1}d^3q_1\cdots\int_{k_n}d^3q_n \delta_D(\mathbf{q}_{1\cdots n})
=
\frac{1}{N^3V_f^{n-1}}
\sum_{[\bfx]}
\prod_{i=1}^n 
{\rm FFTW}
\left(
    \tilde{\iota}_{k_i}[\bfq]
\right)\,.
\end{equation}
where we define $\tilde{\iota}_{k_i}[\bfq]$, 
analogous to $\tilde{I}_{k_i}[\bfq]$ above, 
as the Fourier grid which takes unity within the spherical shell bounded by
$|k - k_i| < \delta k/2$ and zero elsewhere.

\section{Approximate Analytic Number of Polygons} \label{app:analytic}
In the polyspectra estimator, we have the normalization constant
$\frac{V_{\tilde{n}}}{V_f^{n-1}}$, which is the number of $n$-gons that 
satisfy the closing condition (Dirac-delta) for a given set of 
side-length wavenumber $k_i$.
In this appendix, we shall present the analytical approximation to this 
quantity. 

Using the Dirac-delta (\refeq{dirac}), we rewrite the $3n$-dimensional volume 
$V_{\tilde{n}}$ as
\begin{equation}
V_{\tilde{n}} = \int_{k_1}d^3q_1\cdots\int_{k_n}d^3q_n \delta_D(\mathbf{q}_{1\cdots n}) = \frac{1}{(2\pi)^3}\int dx^3\int_{k_1}d^3q_1e^{i\mathbf{x}\cdot\mathbf{q}_1}\cdots \int_{k_n}d^3q_ne^{i\mathbf{x}\cdot\mathbf{q}_n}\,,
\end{equation}
for which we can compute the each $q_i$ integration as
\begin{equation}\label{eq:approx}
\int_k d^3q e^{i\mathbf{q}\cdot \mathbf{x}} = 4\pi \int\limits^{k+\delta k/2}_{k- \delta k/2} dq q^2 \frac{\sin(qx)}{qx} \approx 4\pi \frac{k\sin(kx)}{x}\delta k + \mathcal{O}(\delta k^3).
\end{equation}
Combining the two results, we reduce the number of $n$-gons to one-dimensional
integration as
\begin{equation}\label{eq:volint}
V_{\tilde{n}} = 2^{2n-3}\pi^{n-3} 
\left(\prod_{i=1}^n k_i\delta k\right)
\int d^3x \frac{1}{x^n}
\prod_{i=1}^n \sin\left(k_ix\right)
= 2^{2n-1}\pi^{n-2}
\left(\prod_{i=1}^n k_i\delta k\right)
\int\limits_0^\infty \frac{dx}{x^{n-2}}
\prod_{i=1}^n \sin\left(k_ix\right)\,.
\end{equation}
which is an analytically tractable integral, 
given the assumptions that $k_{i} \geq k_{i+1}$ (to uniquely define the polygon), and the polygon condition that $(\sum_{i\neq j}k_i) \geq k_j$.

In what follows, we shall show the calculations. 
First, we show the general strategy of integrating ${\cal I}_n$ 
in \refapp{gen}, which is followed by the explicit result for 
$n=3$ (\refapp{n3}) and $n=4$ (\refapp{n4}) cases. 
We then present the calculation for the colinear case, 
$k_1=k_2+\cdot+k_n$, where the approximation in \refeq{approx} resulting 
zero for all $n>4$ polygons. We therefore show the explicit expression 
without approximation in \refeq{approx} in \refapp{colinear}.

\subsection{The analytical calculation: general case}\label{app:gen}
To simplify the notation, let us define following function
\be
{\cal I}_n\(k_1,\cdots,k_n\)
\equiv 
\int\limits_0^\infty dx\frac{\prod \sin(k_ix)}{x^{n-2}}
=
\frac12 \int_{-\infty}^\infty 
\frac{dx}{x^{n-2}}\prod_{i=1}^n \sin\(k_ix\),
\ee
where we use that the integrand is an even function of $x$. 
Here, we show the general strategy of solving for the function ${\cal I}_n$.
First, we use the following identity:
\be
\frac{\sin(kx)}{kx}
=
\int_{-1}^1 \frac{d\mu}{2} \, e^{ikx\mu}\,,
\ee
to transform the integration as 
\be
{\cal I}_n(k_1,\cdots,k_n)
=
k_1\cdots k_{n-2}
\int_{-\infty}^\infty \frac{dx}{2}
\int_{-1}^1 \frac{d\mu_1}{2}
\cdots
\int_{-1}^1 \frac{d\mu_{n-2}}{2}
e^{ix\(k_1\mu_1 + \cdots + k_{n-2}\mu_{n-2}\)}
\sin\(k_{n-1}x\)
\sin\(k_{n}x\)\,.
\ee
We then use 
\ba
\sin\(k_{n-1}x\) \sin\(k_{n}x\)
=&
\frac{1}{(2i)^2}
\[e^{ik_{n-1}x} - e^{-ik_{n-1}x}\]
\[e^{ik_{n}x} - e^{-ik_{n}x}\]
\vs
=&
\frac{1}{4}
\(
  e^{i(k_{n-1}-k_n)x} 
+ e^{-i(k_{n-1}-k_n)x} 
- e^{i(k_{n-1}+k_n)x} 
- e^{-i(k_{n-1}+k_n)x}
\)\,,
\ea
to convert the integrations in terms of the Dirac-delta operator:
\be
\delta^D(k) 
= \int_{-\infty}^\infty \frac{dx}{2\pi} e^{ikx}\,. 
\ee
That is, 
\ba
{\cal I}_n(k_1,\cdots,k_n)
=&
\frac{\pi}{4}
k_1\cdots k_{n-2}
\int_{-1}^1 \frac{d\mu_1}{2}
\cdots
\int_{-1}^1 \frac{d\mu_{n-2}}{2}
\vs
&\times \Biggl[
\delta^D\(k_1\mu_1 + \cdots + k_{n-2}\mu_{n-2} + k_{n-1} - k_n\)
+
\delta^D\(k_1\mu_1 + \cdots + k_{n-2}\mu_{n-2} - k_{n-1} + k_n\)
\vs
&\qquad-
\delta^D\(k_1\mu_1 + \cdots + k_{n-2}\mu_{n-2} + k_{n-1} + k_n\)
-
\delta^D\(k_1\mu_1 + \cdots + k_{n-2}\mu_{n-2} - k_{n-1} - k_n\)
\Biggl]
\vs
=&
\frac{\pi}{2}
k_1\cdots k_{n-2}
\int_{-1}^1 \frac{d\mu_1}{2}
\cdots
\int_{-1}^1 \frac{d\mu_{n-2}}{2}
\vs
&\times \Biggl[
\delta^D\(k_1\mu_1 + \cdots + k_{n-2}\mu_{n-2} + k_{n-1} - k_n\)
- \delta^D\(k_1\mu_1 + \cdots + k_{n-2}\mu_{n-2} - k_{n-1} - k_n\)
\Biggl]
\vs
=&
\frac{\pi}{2}
\int_{-k_1}^{k_1} \frac{d\kappa_1}{2}
\cdots
\int_{-k_{n-2}}^{k_{n-2}} \frac{d\kappa_{n-2}}{2}
\Biggl[
\delta^D\(\kappa_1 + \cdots + \kappa_{n-2} + k_{n-1} - k_n\)
- \delta^D\(\kappa_1 + \cdots + \kappa_{n-2} - k_{n-1} - k_n\)
\Biggl]\,.
\label{eq:master}
\ea
Going beyond this result requires the conditions to form polygons. 
For general polygons in 3D there is only one condition, the longest side must be less than or equal to the sum of all the other sides. If we don't want to be bothered by the ordering, we can simply state as following: For any $j\in[1,n]$,
\be
k_j \le 
k_1 + \cdots + k_{j-1} + k_{j+1} + \cdots +k_n
= \sum_{i=1}^n k_i - k_j\,.
\label{eq:polygon}
\ee
With this condition, we can see that there must be a solution for $\mu_i\in(0,1)$ satisfying
\be
k_n = 
k_1\mu_1 + \cdots + k_{n-2}\mu_{n-2} + k_{n-1}\,.
\ee
For later use, we find following identity for the integration of the Dirac delta useful:
\be
\int_{-k}^k d\kappa \delta^D(\kappa-a)
=
\Theta(k-a,k+a).
\ee
Here $\Theta(x_1,x_2,\cdot,x_n)$ is a multi-dimensional Heaviside-Theta function, which is 1 only if none of the $x_i$ are not positive. 

\subsection{The analytical calculation for triangles ($n=3$)}\label{app:n3}
For the bispectrum ($n=3$), \refeq{master} reduces to 
\ba
{\cal I}_3 
=& 
\frac{\pi}{2}
\int_{-k_1}^{k_1} \frac{d\kappa_1}{2}
\[
\delta^D(\kappa_1+k_2-k_3)
-
\delta^D(\kappa_1-k_2-k_3)
\]
\vs
=&
\frac{\pi}{4}
\[
\Theta(k_1+k_2-k_3,k_1-k_2+k_3)
-
\Theta(k_1-k_2-k_3,k_1+k_2+k_3)
\]\,.
\ea
Without loss of generality, let us order the wavenumbers so that 
$k_1\ge k_2\ge k_3$. Then, from the polygon inequality in \refeq{polygon}, the first term becomes
\be
\Theta(k_1+k_2-k_3,k_1-k_2+k_3) = 1\,,
\ee
and the second term is
\be
\Theta(k_1-k_2-k_3,k_1+k_2+k_3)
=
\left\{
\begin{array}{cc}
1/2        &,~k_1=k_2+k_3 \\
0          &,~k_1<k_2+k_3
\end{array}
\right.\,.
\ee
Therefore, 
\be
{\cal I}_3
=
\left\{
\begin{array}{cc}
\pi/8        &,~k_1=k_2+k_3 \\
\pi/4          &,~k_1<k_2+k_3
\end{array}
\right.\,,
\ee
and
\be
V_B = 
\left\{
\begin{array}{cc}
4\pi^2 k_1k_2k_3\(\delta k\)^3 &,~k_1=k_2+k_3 \\
8\pi^2 k_1k_2k_3\(\delta k\)^3 &,~k_1<k_2+k_3
\end{array}
\right.\,.
\ee

\subsection{The analytical calculation for trispectrum}\label{app:n4}
For the trispectrum ($n=4$), \refeq{master} reduces to 
\ba
{\cal I}_4
=& 
\frac{\pi}{2}
\int_{-k_1}^{k_1} \frac{d\kappa_1}{2}
\int_{-k_2}^{k_2} \frac{d\kappa_2}{2}
\[
\delta^D(\kappa_1+\kappa_2+k_3-k_4)
-
\delta^D(\kappa_1+\kappa_2-k_3-k_4)
\]
\vs
=&
\frac{\pi}{4}
\int_{-k_1}^{k_1} \frac{d\kappa_1}{2}
\[
\Theta(k_2+\kappa_1+k_3-k_4,k_2-\kappa_1-k_3+k_4)
-
\Theta(k_2+\kappa_1-k_3-k_4,k_2-\kappa_1+k_3+k_4)
\]\,.
\ea
Again, without loss of generality, we can order the wavenumbers so that $k_1\ge k_2\ge k_3\ge k_4$. 

We first consider the case where $k_1=k_2+k_3+k_4$. The integration then becomes
\ba
{\cal I}_4
=& 
\frac{\pi}{4}
\int_{-k_1}^{k_1} \frac{d\kappa_1}{2}
\[
\Theta(\kappa_1+k_1-2k_4,-\kappa_1+k_1-2k_3)
-
\Theta(\kappa_1+2k_2-k_1,k_1-\kappa_1)
\]
\vs
=&
\frac{\pi}{4}
\[
\int_{{\rm max}(-k_1,-k_1+2k_4)}^{{\rm min}(k_1,k_1-2k_3)} \frac{d\kappa_1}{2}
-
\int_{{\rm max}(-k_1,k_1-2k_2)}^{k_1} \frac{d\kappa_1}{2}
\]
=
\frac{\pi}{8}
\[
k_1-2k_3 - (-k_1+2k_4) - (k_1 - (k_1-2k_2))
\]
=
0\,.
\ea
That is, the number of trispectrum using \refeq{approx} vanishes for the 
colinear quadraliterals. We present the expression without using the 
approximation in \refapp{colinear}.

For the other cases, $k_1<k_2+k_3+k_4$, we integrate ${\cal I}_4$ as following. The first Heaviside-Theta function vanishes unless,
\be
-k_2-k_3+k_4 \le \kappa_1 \le k_2-k_3+k_4,
\ee
so that we integrate the first term as
\ba
\int_{-k_1}^{k_1} \frac{d\kappa_1}{2}
\Theta(k_2+\kappa_1+k_3-k_4,k_2-\kappa_1-k_3+k_4)
=&
\int_{{\rm max}(-k_1,-k_2-k_3+k_4)}^{{\rm min}(k_1,k_2-k_3+k_4)} \frac{d\kappa_1}{2}
\vs
=&
\frac12
\[
{\rm min}(k_1,k_2-k_3+k_4)
-
{\rm max}(-k_1,-k_2-k_3+k_4)
\]\,.
\ea
From \refeq{polygon}, we know that $k_1>-k_2-k_3+k_4$, and $-k_1<k_2-k_3+k_4$, and, of course, $k_2-k_3+k_4>-k_2-k_3+k_4$. Therefore,
\ba
&\frac12
\[
{\rm min}(k_1,k_2-k_3+k_4)
-
{\rm max}(-k_1,-k_2-k_3+k_4)
\]
\vs
=
&
\frac12\[
2k_2
+
\(k_1 - k_2 + k_3 - k_4\)
\Theta\(k_2 - k_3 + k_4 - k_1\)
+
\(k_1 - k_2 - k_3 + k_4\)
\Theta\(-k_1 + k_2 +k_3 - k_4\)
\]\,.
\ea
The second Heaviside-Theta function vanishes unless,
\be
-k_2+k_3+k_4 \le \kappa_1 \le k_2+k_3+k_4.
\ee
Using that, we find, from \refeq{polygon}, $k_1< k_2+k_2+k_4$ and $-k_1<-k_2+k_3+k_4$, and of course, $-k_2+k_3+k_4< k_2+k_3+k_4$. Therefore,
\ba
\int_{-k_1}^{k_1} \frac{d\kappa_1}{2}
\Theta(k_2+\kappa_1-k_3-k_4,k_2-\kappa_1+k_3+k_4)
=&
\int_{{\rm max}(-k_1,-k_2+k_3+k_4)}^{{\rm min}(k_1,k_2+k_3+k_4)} \frac{d\kappa_1}{2}
=
\int_{-k_2+k_3+k_4}^{k_1}
\frac{d\kappa_1}{2}
\vs
=&
\(\frac{k_1+k_2-k_3-k_4}{2}\)
\Theta(k_1+k_2-k_3-k_4)\,.
\ea
Combining all, we find that
\ba
I_4 =&
\frac{\pi}{8}
\left[
2k_2
+
\(k_1 - k_2 + k_3 - k_4\)
\Theta\(k_2 - k_3 + k_4 - k_1\)
\right.
\vs
&
\left.
\quad+
\(k_1 - k_2 - k_3 + k_4\)
\Theta\(-k_1 + k_2 +k_3 - k_4\)
+
\(k_1+k_2-k_3-k_4\)
\Theta(k_1+k_2-k_3-k_4)
\right]\,.
\ea
We then calculate the $V_T$ as 
\be
V_T = 8\pi^3k_1k_2k_3k_4\delta k^4(-k_1+k_2+k_3+3k_4 - |k_1-k_2-k_3+k_4|)\,.
\ee

\subsection{Exact analytic colinear case} \label{app:colinear}
For the higher-order polyspectra, the lowest order theory predicts zero polygons for the colinear case $k_1 = \sum_{2}^nk_i$ but this is clearly not the case through either direct summation tests or estimation through our method. Simply adding a few higher-order terms in the series of \refeq{approx} causes the integration to diverge so we instead need to solve the exact case directly. 

The full calculation for the integration that we approximate in 
\refeq{approx} yields
\begin{equation}
\int\limits_{k-\delta k/2}^{k+\delta k/2}dqq^2\frac{\sin(qx)}{qx} = \frac{-x \delta k \cos(kx) \cos(x \delta k/2) + 
 2 \left[\cos(kx) + kx \sin(kx)\right] \sin(x \delta k/2)}{x^3}\,.
 \label{eq:iota_exact}
\end{equation}
Restricting to the colinear case, we calculate the Fourier volume integration
for the bispectrum:
\begin{equation}
\label{eq:bicol}
V_B = \frac{32\pi^2}{6144}\delta k^3
\left[104k_3^2\delta k + 5\delta k^3 + 8k_2^2\left(96k_3 + 13 \delta k \right) + 8k_2k_3\left(96k_3 + 13\delta k\right) \right]\,,
\end{equation}
and for the trispectrum
\begin{equation}\label{eq:tricol}
V_{\tilde{T}} =
\left\{
\begin{array}{lc}
\frac{128\pi^3\delta k^5}{725760}
\left[
\delta k\left\{468k_4^2\delta k + 17\delta k^3 + 36k_3^2(105k_4 + 13\delta k) + 36k_3k_4(105k_4 + 13\delta k)\right\} \right.& \\ 
\qquad+ 36k_2^2\left\{21k_3(28k_4+5\delta k) + \delta k(105k_4 + 13\delta k)\right\} &  k_4 > \delta k\\
\qquad\left.+ 36 k_2\left\{21k_3^2(28k_4+5\delta k) + k_4\delta k(105k_4 + 13\delta k) + k_3(588k_4^2 + 210k_4\delta k + 13\delta k^2)\right\}\right] & \\
\frac{128\pi^3\delta k^6}{725760}
\left[24948k_2^2k_3 + 24948k_2k_3^2 + 36(118k_2^2 + 811k_2k_3 + 118 k_3^2)\delta k + 4248(k_2+k_3)\delta k^2 + 485 \delta k^3\right] & k_4=\delta k \\
\frac{128\pi^3\delta k^7}{725760}(29196k_2^2 + 58392k_2\delta k + 8981\delta k^2) & k_3=\delta k \\
\frac{12364288\pi^3\delta k^9}{725760} & k_2=\delta k
\end{array}
\right.\,,
\end{equation}
by using the {\sf Mathematica} package.

For the Bispectrum case \refeq{bicol} differs only about 1\% from the 
lowest order estimation. For trispectrum, where the lowest order approximation
predicts zero polygons, \refeq{tricol} dramatically improves the match 
between the numerical calculation and analytical estimation, as we show in
\reffig{triflat}.

\end{document}